\documentclass[article,11pt,letterpaper]{article}
\usepackage{fullpage}
\usepackage[utf8]{inputenc}

\usepackage{amsmath, amssymb, amsfonts, amsthm}

\usepackage{natbib}
\usepackage{url}
\usepackage[colorlinks,linkcolor=blue,citecolor=blue,urlcolor=blue]{hyperref}

\usepackage[T1]{fontenc}
\usepackage{authblk}
\usepackage[usenames,dvipsnames]{xcolor}
\usepackage{verbatim, color}
\usepackage{afterpage}
\usepackage{multirow}
\usepackage{graphicx}
\usepackage{algorithm}
\usepackage{placeins}
\usepackage{algpseudocode}
\newcommand{\iid}{\buildrel iid \over\sim}
\usepackage{grffile}

\theoremstyle{definition}

\newcommand{\Tcal}{\mathcal{T}}

\def\bs{\boldsymbol{s}}
\def\bt{\boldsymbol{t}}
\def\bu{\boldsymbol{u}}
\def\bv{\boldsymbol{v}}

\def\bx{\boldsymbol{x}}
\def\by{\boldsymbol{y}}
\def\bz{\boldsymbol{z}}

\def\bC{\boldsymbol{C}}

\def\bE{\boldsymbol{E}}

\def\bH{\boldsymbol{H}}
\def\bI{\boldsymbol{I}}

\def\bS{\boldsymbol{S}}
\def\bT{\boldsymbol{T}}

\def\bV{\boldsymbol{V}}

\def\bZ{\boldsymbol{Z}}

\def\zeros{\boldsymbol{0}}

\def\bdelta{\boldsymbol{\delta}}
\def\bgamma{\boldsymbol{\gamma}}
\def\bmu{\boldsymbol{\mu}}
\def\btheta{\boldsymbol{\theta}}

\def\bomega{\boldsymbol{\omega}}

\def\bDelta{\boldsymbol{\Delta}}

\def\bPi{\boldsymbol{\Pi}}
\def\bPsi{\boldsymbol{\Psi}}

\def\bOmega{\boldsymbol{\Omega}}
\def\bSigma{\boldsymbol{\Sigma}}

\def\diag{\textup{diag}}

\def\R{\mathbb{R}}
\def\E{\textup{E}}
\def\var{\text{Var}}

\newcommand{\PP}{\mathbb{P}} 
\DeclareMathOperator*{\cov}{Cov}

\DeclareMathOperator*{\vect}{vec}

\def\sigsq{\sigma^2}

\def\N{\textup{N}}

\def\TN{\textup{TN}}
\def\IG{\textup{IG}}
\def\Exp{ \textup{Exp} }
\def\Bern{ \textup{Bernoulli} }

\newcommand{\bpm}{\begin{pmatrix}}
\newcommand{\epm}{\end{pmatrix}}

\newcommand{\bfl}[1]{\begin{flalign} #1 \end{flalign}}
\newcommand{\bfln}[1]{\begin{flalign*} #1 \end{flalign*}}

\usepackage[textwidth=2cm, textsize=scriptsize]{todonotes}

\def\namedlabel#1#2{\begingroup
    #2%
    \def\@currentlabel{#2}%
    \phantomsection\label{#1}\endgroup
}

\def\moracle{{\textup{Oracle}} }
\def\mdist{{\textup{Dist}}}
\def\mphylo{{\textup{PhyloBCG}}}

\usepackage{pgf}
\usepackage{tikz}
\usetikzlibrary{arrows,automata}
\usepackage{tikz/tikzit}

\tikzstyle{rectangle}=[fill=white, draw=black, shape=rectangle, align=center]
\tikzstyle{leftAligned}=[fill=white, draw=white, shape=rectangle, anchor=base west]
\tikzstyle{diamond}=[fill=white, draw=black, shape=diamond, align=center]
\tikzstyle{ellipse}=[fill=white, draw=black, shape=ellipse, align=center]
\tikzstyle{circle}=[fill=white, draw=black, shape=circle]
\tikzstyle{nobound}=[fill=white, draw=none, shape=rectangle, align=center]
\tikzstyle{dashedrectangle}=[fill=white, draw=black, shape=rectangle, align=center, dashed]

\tikzstyle{arrow->}=[->, line width=0.8pt]
\tikzstyle{new edge style 1}=[-, line width=0.8pt, draw=black]
\tikzstyle{dashed line}=[-, draw=red, densely dashed]
\tikzstyle{dashedArrow}=[->, line width=0.5pt, dashed=]

\graphicspath{{./figures/}}
\usepackage{grffile}

\begin{document}

\title{Phylogenetically informed Bayesian truncated copula graphical models for microbial association networks}
\author[]{Hee Cheol Chung, Irina Gaynanova, Yang Ni} 
\affil[]{\small Department of Statistics, Texas A\&M University, College Station, TX 77843, USA}
\date{}
\maketitle
\begin{abstract}
Microorganisms play a critical role in host health. The advancement of high-throughput sequencing technology provides opportunities for a deeper understanding of microbial interactions. However, due to the limitations of 16S ribosomal RNA sequencing, microbiome data are zero-inflated, and a quantitative comparison of microbial abundances cannot be made across subjects. By leveraging a recent microbiome profiling technique that quantifies 16S ribosomal RNA microbial counts, we propose a novel Bayesian graphical model that incorporates microorganisms' evolutionary history through a phylogenetic tree prior and explicitly accounts for zero-inflation using the truncated Gaussian copula. Our simulation study reveals that the evolutionary information substantially improves the network estimation accuracy. We apply the proposed model to the quantitative gut microbiome data of 106 healthy subjects, and identify three distinct microbial communities that are not determined by existing microbial network estimation models. We further find that these communities are discriminated based on microorganisms' ability to utilize oxygen as an energy source.
\end{abstract}


\section{Introduction}
\subsection{Microbial association network and its importance}
The gut is one of the most significant habitats of a myriad of microbial communities that play critical roles in their host's health. Well-balanced gut microbial communities provide many health benefits, such as maintaining metabolic homeostasis and high functioning immune system \citep{martinez2016gut,kim2016gut,cani2019microbial}. The imbalance of the gut microbiome (dysbiosis) has been related to a variety of human diseases
\citep{cho2012human,lynch2016human}. 
The gut microbial balance is maintained by complex microbial interactions such as metabolites consumption, production, and exchange. Microbiome dysbiosis occurs when these interactions are interrupted by environmental alterations such as diet change, antibiotic consumption, and chemical exposure. These changes may deplete nutrients for beneficial microbes and create favorable surroundings for disease-causing bacteria to flourish. Nevertheless, some microbes help the microbial communities to maintain their stability under the environmental changes by providing energy sources and necessary metabolites \citep{zhang2019facing}. Because of the complexity of the functional roles of microbes, identifying microbial association networks, that is, microbe-microbe interaction networks, is crucial for fundamental understanding of the gut microbiome,  a key contributor to the host's health.


\subsection{Motivating application: quantitative microbiome profiling data}
 Microbiome data collected from 16S ribosomal RNA (rRNA) sequencing are compositional in that each subject has an arbitrary total microbial count determined by the sequencing instrument \citep{gloor2017microbiome}. Hence, a quantitative comparison of microbial abundances cannot be made across subjects as only the information on relative abundances within a subject are available from such data.
Furthermore, compositional data raise a concern for biased estimates of association since a change in absolute abundance of one microbe affects the relative abundance of all the microbes \citep{vandeputte2017quantitative}. The recently developed quantitative microbiome profiling (QMP) techniques account for these compositional limitations by adjusting microbial counts from 16S rRNA sequencing using cell counts and sequencing depths. In this work, we utilize this recent development by considering the QMP data of \citet{vandeputte2017quantitative} with $n=106$ healthy subjects' gut microbiome. We focus on estimating genus-level association networks with the aim of understanding the overall configurations of healthy gut microbial communities and their interactions.

\subsection{Graphical models and network estimation}
Gaussian graphical model is a popular tool for modeling an association network via an undirected graph, where an edge between the two nodes represents the conditional dependence, and an absence of an edge represents conditional independence. Under the Gaussian assumption, this graph structure is fully encoded in the concentration matrix (the inverse covariance matrix) as a zero off-diagonal entry is equivalent to the conditional independence between the corresponding variables. Thus, multiple methods focus on sparse estimation of concentration matrices. Neighborhood selection \citep{meinshausen2006high} recovers the sparse graph structure by performing $L_{1}$-regularized regression of each node on the rest. \citet{yuan2007model,banerjee2008model,dahl2008covariance,friedman2008sparse} directly estimate the sparse concentration matrix by optimizing the $L_{1}$-penalized log-likelihood function, the so-called graphical lasso. \citet{wang2012bayesian} propose a Bayesian counterpart of the graphical lasso using the Laplace prior on the off-diagonal elements of the concentration matrix. \citet{roverato2002hyper,dobra2011bayesian,lenkoski2011computational} consider a G-Wishart prior for the concentration matrix, of which the posterior inference is computationally more expensive than \cite{wang2012bayesian}. For better scalability, \citet{wang2015scaling} develop a continuous spike-and-slab prior for the off-diagonals of the concentration matrix. Furthermore, the Gaussian graphical models can be extended to non-Gaussian data via latent Gaussian copula models. \citet{liu2012high} consider Gaussian copula model for skewed continuous distributions. \citet{fan2017high} consider extension to mixed binary-continuous variables via latent Gaussian copula. \citet{dobra2011copula} consider a Bayesian latent Gaussian copula for graph estimation with binary and ordinal variables, where they approximate the likelihood function using the extended rank likelihood \citep{hoff2007extending}. 

Despite the significant advancements in Gaussian graphical models, they are not appropriate for estimating microbial association networks. Microbiome data obtained from high-throughput sequencing are heavily right skewed and zero-inflated. The zeros, furthermore, are not necessarily absolute, but are often due to the limited sequencing depth. Thus, direct application of Gaussian graphical models to zero-inflated sequencing data leads to inaccurate estimation and inference. To address these challenges, several graphical models for zero-inflated data have been proposed. \citet{osborne2020latent} model microbial counts using Dirichlet-multinomial distribution with a latent Gaussian graphical model. \citet{zhou2020identification} consider a zero-inflated latent Ising model for microbial association network estimation. \citet{mcdavid2019graphical} propose a multivariate hurdle model, which is a mixture of degenerate (at 0) and Gaussian distributions. SPIEC-EASI \citep{kurtz2015sparse} is a two-stage inference procedure specifically designed for compositional microbiome data. \citet{yoon2019spring} propose Semi-Parametric Rank-based approach for INference in Graphical model (SPRING) based on truncated latent Gaussian copula \citep{yoon2020sparse}. \citet{ma2020joint} proposes truncated Gaussian graphical model.




 \subsection{The major limitation of existing network estimation models and our proposal}
The aforementioned microbial network estimation models share a common limitation: they do not take advantage of additionally available evolutionary information for reverse-engineering the graph structure. 
The information on microbes' genetic similarities is available in a form of a phylogenetic tree, however to our knowledge the phylogenetic tree is not taken into account by existing methods for estimation of microbial networks.  Since microbial interactions, positive (e.g., mutualism) or negative (e.g., competition), increase with the increase in microbes' genetic similarity \citep{rohr2014, peralta2016merging}, evolutionary information encoded in a phylogenetic tree has great potential in improving the accuracy of microbial associations network estimation.

In this work, we propose a Bayesian truncated Gaussian copula graphical model for microbial association networks that takes advantage of available evolutionary information. Our major contributions are three-fold. First, we provide a general framework for incorporating evolutionary history into the estimation of microbe-microbe association networks. We model the phylogenetic tree as a Gaussian diffusion process in the latent space, 
which allows us to represent the microbes and their ancestors as (correlated) Gaussian vectors.  Our framework is not limited to the phylogenetic tree and can accommodate any prior knowledge that is expressed in a tree, e.g., a taxonomic rank tree. We formulate the prior probability model on graph so that the microbes that are closer to each other on the tree have a higher edge inclusion probability. Our simulation study reveals that our approach significantly improves the graph estimation accuracy compared to the methods that do not take advantage of the tree structure (Section~\ref{sec:simulation}). Second, the proposed model effectively handles zero-inflation resulting from limited sequencing depth. We consider the observed zeros as truncated realizations of unobserved random quantities that are below certain thresholds. In particular, we establish a Bayesian formulation of the truncated Gaussian copula model \citep{yoon2020sparse} and develop an efficient Gibbs sampling algorithm. Third, the proposed approach facilitates the statistical inference on the estimated network. For each pair of nodes, an edge connectivity is immediately available from the posterior sample, which provides a convenient way to control the posterior expected FDR \citep{mitra2013bayesian,peterson2015bayesian}. Finally, while our model is designed for quantitative microbiome data, it can also be applied to compositional data using modified central log ratio transformation \citep{yoon2019spring}.

The rest of this paper is organized as follows. In Section \ref{sec:model}, we introduce the proposed graphical model. In Section \ref{sec:inference}, we discuss posterior inference. In Section \ref{sec:simulation}, we evaluate the graph estimation accuracy of the proposed model on simulated datasets. In Section \ref{sec:application}, we analyze quantitative microbiome profiling data of \citet{vandeputte2017quantitative}, and compare our results to SPRING \citep{yoon2019spring} and SPIEC-EASI \citep{kurtz2015sparse}.

\section{Bayesian truncated Gaussian copula graphical model}\label{sec:model}

\begin{figure}[t]
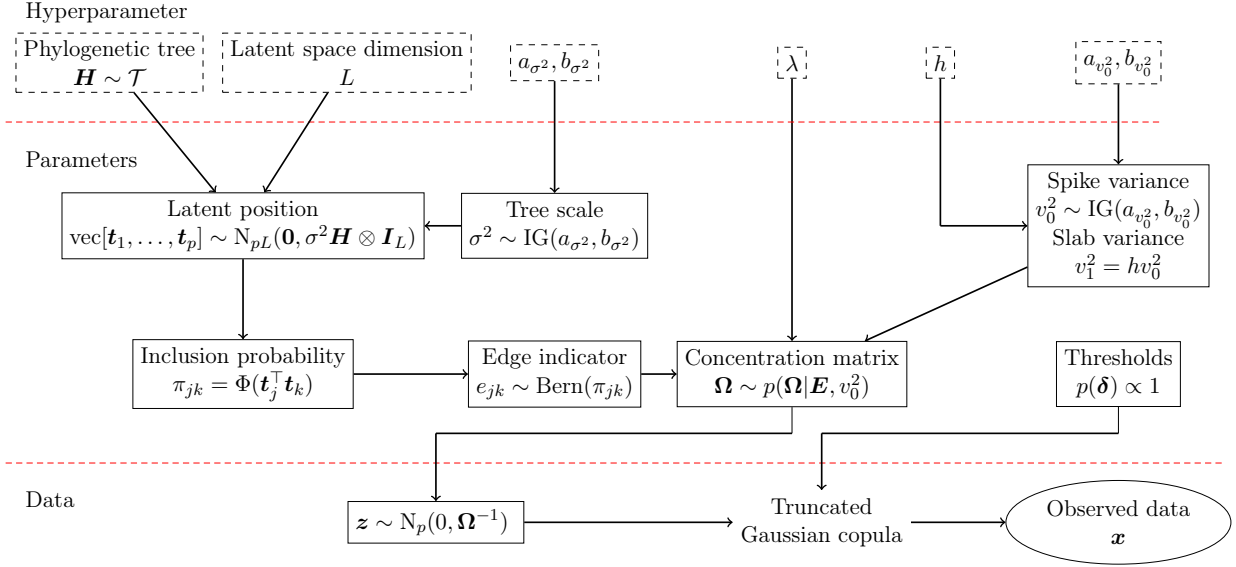

		\ctikzfig{tikz/BCG_scheme}
		\caption{Schematic illustration of the phylogenetically informed Bayesian truncated Gaussian copula graphical model. Hyperparameters that are held constant are given in boxes with dashed-line. The quantities that need posterior inference are illustrated in boxes with solid-line.  The ellipse with solid line represents observed data. }
		\label{fig:schema}
\end{figure}

In Section \ref{ssec:copula}, we discuss the semiparametric modeling of conditional dependencies for zero-inflated data through a truncated Gaussian copula model with a sparse concentration matrix. In Section \ref{ssec:tree}, we introduce a prior model that incorporates the phylogenetic tree to facilitate posterior inference of microbial associations. The complete hierarchical model is summarized in Figure \ref{fig:schema}.

\subsection{Truncated Gaussian copula graphical model}\label{ssec:copula}
Let $\bx=(x_{1},\ldots,x_{p})^{\top}$ denote the zero-inflated abundances of $p$ microbes. This is either directly the counts resulting from quantitative microbiome profiling, e.g. motivating data from \citet{vandeputte2017quantitative}, or transformed
compositional microbiome data via the modified central log-ratio transformation \citep{yoon2019spring}. 
We propose to model $\bx$ such that only the microbial abundances that are larger than certain thresholds can be observed. Specifically, we assume that there exist latent $\bx^{*}=(x_{1}^{*},\ldots,x_{p}^{*})^{\top}$ representing the true abundances such that
\begin{flalign}\label{eq:truncation}
x_{j} &= 1(x_{j}^{*}>c_{j})x_{j}^{*}, \quad j=1,\ldots,p,
\end{flalign}
where $1(\cdot)$ is the indicator function and $c_{j}$ is the unknown threshold for the $j$th variable. 
We call $x_j$ a \textit{truncated} variable if $x_{j}^{*}$ is less than $c_j$, and an \textit{observed} variable, otherwise. Let $F_{j}$ be the marginal cumulative distribution function (cdf) of the $j$th latent variable $x_{j}^{*}$,  $\Phi$ be the cdf of standard Gaussian, and $f_{j}=\Phi^{-1}\circ F_{j}$, where  we assume that $F_{j}$'s are continuous. 
The truncated Gaussian copula model \citep{yoon2020sparse} assumes
\bfl{\label{eq:motr}
   & z_{j} = f_{j} (x_{j}^{*}), \quad j=1\ldots,p,\\ \label{eq:copula}
	&\bz=(z_{1},\ldots,z_{p})^{\top} \sim \N_{p}(\zeros,\bOmega^{-1}),
}
where $\bOmega^{-1} \succ 0$ is the positive definite correlation matrix.

Since $f_{j}$'s are monotone continuous, we can write \eqref{eq:truncation} as $x_{ij}=1\{f_{j}(x_{ij}^{*})>f_{j}(c_{j}) \}x_{ij}^{*}$ $=1(z_j>\delta_j )x_{ij}^{*}$, where $\delta_j=f_j(c_j)$. We denote the truncated and the observed sub-vectors of $\bx$ by $\bx_{t}\in \R^{p_{t}}$ and $\bx_{o} \in \R^{p_{o}}$, respectively,  where $p_{t}+p_{o}=p$. Likewise, let $\bz_{t}$ and $\bz_{o}$ be the corresponding latent Gaussian vectors, and let $\bdelta_{t}$ and $\bdelta_{o}$ be their thresholds. 
Given the thresholds, the conditional distribution of $\bz$ is given by
\begin{flalign}\label{eq:density}
	p(\bz_{o},\bz_{t}|\bz_{o}>\bdelta_{o},\bz_{t}<\bdelta_{t},\bOmega)= 
	\frac{\N_{p}(\bz_{o},\bz_{t}|\zeros,\bOmega^{-1}) }{\PP(\bz_{o}>\bdelta_{o},\bz_{t}<\bdelta_{t}|\bdelta,\bOmega)}  1(\bz_{o}>\bdelta_{o}) 1(\bz_{t}<\bdelta_{t}).
\end{flalign}
Unlike the approach in \citet{dobra2011copula} that only uses the relative ranks of the observed data, we condition on the observed value of $\bz_{o}$, which subsequently allows us to sample truncated variables $\bz_{t}$ from the posterior distribution as discussed in Section \ref{sec:inference}.

Because of the multivariate normality of $\bz$, zero entries of $\bOmega=[\omega_{jk}]_{1\leq j, k\leq p}$ imply the conditional indepdendence between the corresponding variables. The dependency structure of $\bz$ can be graphically summarized as an undirected graph $G=(\bz,\bE)$ with an adjacency matrix $\bE=[e_{jk}]_{1\leq j, k\leq p}$, where nodes $z_j$ and $z_k$ are connected (denoted by $e_{jk}=1$) if $\omega_{jk}\neq0$.
 Consequently, learning the graph structure (adjacency matrix) $\bE$ is equivalent to finding the sparse pattern of $\bOmega$. To encourage sparsity, we follow a similar strategy as in \cite{wang2015scaling} by assigning a spike-and-slab prior on the off-diagonal elements of $\bOmega$ and an exponential prior on the diagonal elements,
\bfl{
\label{eq:SSSL2}
p(\bOmega|\bE,v_{0}^2) =  C(\bE,v_{0}^2)^{-1} 
1(\bOmega\succ 0) & \prod_{j<k} \Big\{ (1-e_{jk})\N(\omega_{jk}|0,v_{0}^2) + e_{jk}\N(\omega_{jk}|0,hv_{0}^2) \Big\} \\
\nonumber
 \times & \prod_{j=1}^{p} \Exp(\omega_{jj}|\frac{\lambda}{2}), 
}
where  $\Exp(\cdot|\lambda)$ is the exponential density function with rate parameter $\lambda$,  $v_{0}^{2}$ is the spike variance, $h\gg1$ is a large constant such that the slab variance $hv_{0}^{2}\gg v_{0}^{2}$, and $C(\bE,v_{0}^2)$ is the normalizing constant. 

We impose  an improper uniform prior on the thresholds $\bdelta=(\delta_{1},\ldots,\delta_{p})^{\top}\sim\pi(\bdelta)\propto 1$, a conjugate inverse-gamma prior on the spike variance $v_{0}^2 \sim \IG(a_{v_{0}},b_{v_{0}})$, and Bernoulli-like priors on the edge indicators
\bfl{
\label{eq:eprior}
p(\bE)\propto C(\bE,v_{0}^2)\prod_{j<k} \mbox{Ber}(e_{jk}|\pi_{jk}).
} 
Including the normalizing constant $C(\bE,v_{0}^2)$ in \eqref{eq:eprior} serves to cancel out that in \eqref{eq:SSSL2}, facilitating the posterior computation in updating $\bE$ \citep{wang2015scaling}.

\subsection{Incorporating phylogenetic tree}\label{ssec:tree}

Evolution plays an important role in shaping the interaction patterns of microbes  \citep{peralta2016merging}.  We will exploit the evolution footprints in identifying microbial association networks through a novel phylogenetic tree prior.  The proposed prior is a  distribution on edge inclusion probabilities $\bPi=[\pi_{jk}]_{1\leq j, k\leq p}$  that encourages the interactions, positive (e.g., mutualism) or negative (e.g., competition), of phylogenetically similar microbes as they tend to be phenotypically/functionally correlated \citep{martiny2015microbiomes,xiao2018predictive,zhou2021bayesian}. The prior is constructed by first embedding the network in $L$-dimensional Euclidean space through the latent position model \citep{hoff2002latent}, and then arranging the latent positions according to the phylogenetic tree.

\textbf{Latent position model.}
We introduce a latent position $\bt_j=(t_{1j},\dots,t_{Lj})^{\top}\in \mathbb{R}^L$ for each node $z_j$,  and link it to the edge inclusion probability $\pi_{jk}$ through a probit link function
\bfln{ 
    \pi_{jk} = \Phi( \bt_{j}^{\top} \bt_{k} ), \quad j<k.
}
The inner product $\bt_{j}^{\top} \bt_{k}$ measures the similarity between $\bt_{j}$ and $\bt_{k}$, with larger inner product leading to higher prior inclusion probability. We assign a prior on $\bt_j$'s to encourage the interactions between phylogenetically similar microbes.

\begin{figure}[!t]
	\centering
	\includegraphics[width=1\textwidth]{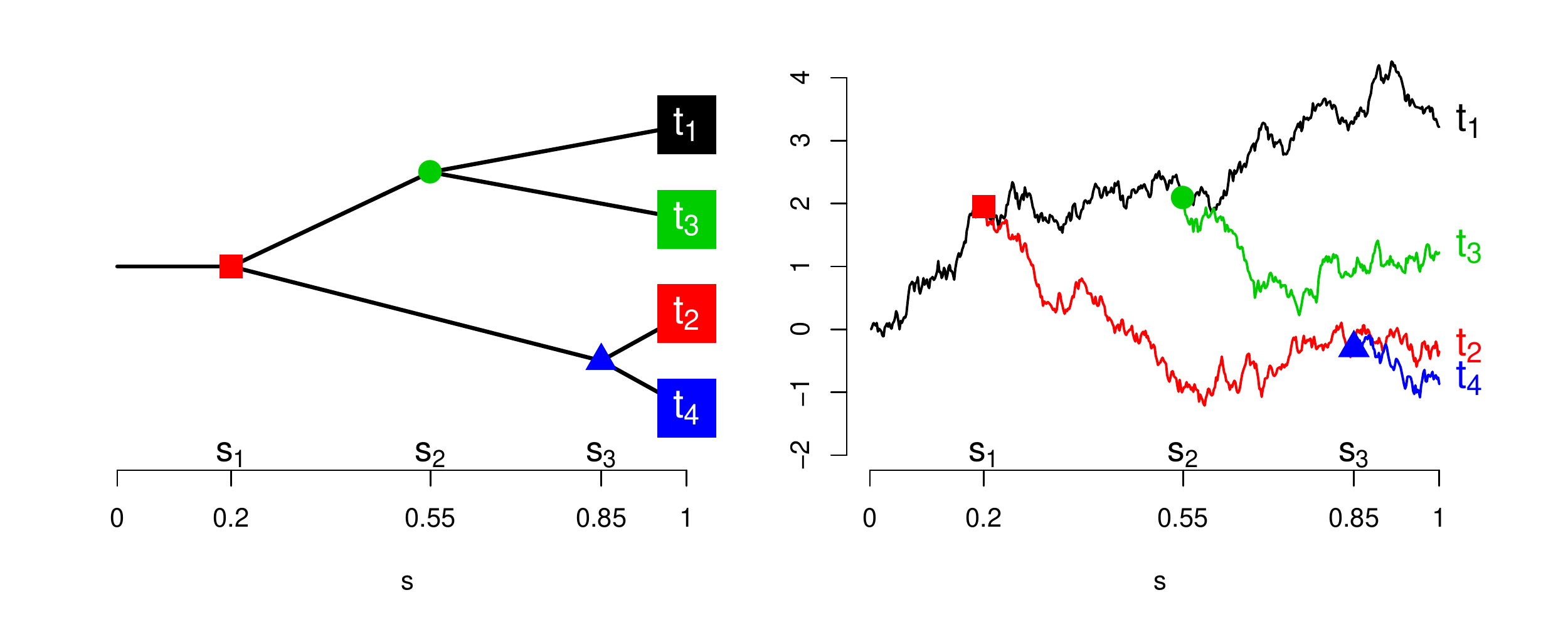}\\
	\caption{An illustrative example of a phylgenetic tree with $p=4$ microbes (left) and the corresponding diffusion process in $\R^{1}$ with $\sigsq=3$ (right). The new branch $t_2$ is split from $t_1$ at the first divergence time $s_1=0.2$. Then new branches $t_3$ and $t_4$ are split from $t_1$ and $t_2$ at the divergence time $s_2=0.55$ and $s_3=0.85$, respectively. The green circle and blue triangle are the most common ancestors of the pairs $(t_1,t_3)$ and $(t_2,t_4)$, whose heights $s_2$ and $s_3$ are the correlations of the pairs, respectively. } \label{fig:diffusiontree}
\end{figure}

\textbf{Phylogenetic tree.} 
Let $\bT=[\bt_{1},\ldots,\bt_{p}]\in \mathbb{R}^{L\times p}$ and let $\bt^\ell=(t_{1\ell},\dots,t_{p\ell})$ be the $\ell$th row of $\bT$. We assume $\bt^{\ell}\iid \N_{p}(\zeros,\sigma^2\bH)$ for $\ell=1,\dots,L$, where $\bH$ is a correlation matrix that reflects the phylogenetic similarity. Our specific choice of $\bH$ is motivated by the following diffusion process. Let $\Tcal$ be a phylogenetic tree with terminal nodes representing the $p$ microbes under investigation and internal nodes representing their common ancestors. Starting from time 0 at the origin (root), the first branch $\bt_{1}$ follows a Brownian motion with variance $\sigsq$ until the divergence time $s_1 \in [0,1]$. Then it splits into two branches, $\bt_{1}$ and $\bt_{2}$, each following the same Brownian motion independently before they split at times $s_2,s_3\in [0,1]$ resulting in $\bt_{3}$ and $\bt_{4}$. This process repeats until the $p$ terminal nodes are reached at time 1. An illustrative example of the diffusion process is provided in Figure \ref{fig:diffusiontree}.

This diffusion process defines a centered multivariate Gaussian distribution on the terminal nodes of $\Tcal$ with covariance matrix $\sigsq \bH$. We define the \textit{height} of each node (split for internal nodes) as its distance from the root (time from 0). The correlation of two terminal nodes equals the height of their most recent common ancestor, which is large for a phylogenetically similar microbes. This multivariate Gaussian prior, together with the latent position model, achieves the desired prior distribution of $\pi_{jk}$ that encourages interactions between phylogenetically similar microbes.
Lastly, we assign a conjugate inverse-gamma prior for the variance parameter $\sigsq\sim \IG(a_{\sigsq},b_{\sigsq})$.

\section{Posterior inference}\label{sec:inference}
The proposed model is parameterized by $\{\bz,\bdelta, \bOmega, \bE,v_{0}^2,\bT,\sigma^2\}$ of which the posterior distribution is not available in closed form.  
We use a Markov chain Monte Carlo (MCMC) algorithm to draw posterior samples from the intractable posterior distribution. Section \ref{ssec:samplezt} discusses Gibbs steps for sampling $\bz$ and $\bdelta$ from their full conditional distributions. Section \ref{ssec:samplelatent} describes Gibbs steps for $\bT$ and $\sigsq$. In Section \ref{supp:spikeandslab} of the Supplemenatry Materials, we provide the Gibbs steps for $\bOmega, \bE,v_{0}^2$ by following \cite{wang2015scaling}. 

\subsection{Full conditionals of truncated observations and thresholds} \label{ssec:samplezt}
Let $\bx_{1},\ldots,\bx_{n} \in \R^{p}$ be a sample from the model \eqref{eq:truncation}--\eqref{eq:copula} and let $\bz_{1},\ldots,\bz_{n} \in \R^{p}$ be the corresponding latent variables. As before, we use subscripts $t$ and $o$, respectively, to denote the truncated and observed sub-vectors (e.g., $\bx_{i,t}$ and $\bx_{i,o}$ are the truncated and observed sub-vectors of $\bx_{i}$).

For observed $\bx_{i,o}$, the corresponding Gaussian variables are defined as $z_{ij,o} =f_j(x_{ij,o})=\Phi^{-1}\circ F_{j}(x_{ij,o})$. A natural estimator for the unknown function $F_{j}$ is the scaled empirical cdf $\widehat{F}_{j}(c)=\{n/(n+1)\}\sum_{i=1}^{n} n^{-1}1(x_{ij}\leq c)$ \citep{klaassen1997efficient}, where the constant term $n/(n+1)$ is needed to make $\Phi^{-1}$ finite. Thus, the estimator of $f_j$ is given by $\Phi^{-1}\circ\widehat F_j$, and we  set $\widehat{z}_{ij,o}=\Phi^{-1}\circ \widehat{F}_{j}(x_{ij,o})$. An alternative approach to estimate $f_j$ using B-spline basis functions has been considered in \citet{mulgrave2020bayesian}; however, we use the empirical cdf for computational efficiency.

Given $\bz_{i,o}=\widehat{\bz}_{i,o}$, we sample $\bz_{i,t}$ from a truncated multivariate Gaussian distribution derived from \eqref{eq:density},
\bfln{
	p(\bz_{i,t}|\widehat{\bz}_{i,o},\bz_{i,t}<\bdelta_{t},\bOmega)= 
	\frac{\N_{p_{i,t}}(\bz_{i,t}|\bmu_{i},\bDelta_{i}) }{\PP(\bz_{i,t}<\bdelta_{t}|\widehat{\bz}_{i,o},\bOmega,\bdelta)}  1(\bz_{i,t}<\bdelta_{i,t}),
}
where $p_{i,t}$ is the number of truncated variables of $\bx_{i}$, and $\bmu_{i}$, $\bDelta_{i}$ are the mean and covariance matrix of $\bz_{i,t}$ given $\bz_{i,o}=\widehat{\bz}_{i,o}$ (detailed expressions are provided in Section \ref{supp:samplezt} of the Supplementary Materials).
The conditional pdf of $\bdelta$ given ${\bz}_{1},\ldots,\bz_{n}$ is proportional to the $n$-product of indicator functions of \eqref{eq:density}. That is, we have the independent uniform full conditional distributions of $\delta_{j}$'s as 
\bfln{
    p(\bdelta| {\bz}_{1},\ldots,{\bz}_{n},\bOmega) &\propto
    \prod_{i=1}^{n} 
    \N_{p}({\bz}_{i,t},\widehat{\bz}_{i,o},|\bOmega)
     1({\bz}_{i,t}<\bdelta_{i,t}) 1(\widehat{\bz}_{i,o}>\bdelta_{i,o})  ,\\
    &\propto \prod_{j=1}^{p} 1({z}_{j,t}^{\max}<\delta_{j}< \widehat{z}_{j,o}^{\min} ),
}
where ${z}_{j,t}^{\max}=\max_{i} {z}_{ij,t}$ and $\widehat{z}_{j,o}^{\min}=\min_{i} \widehat{z}_{ij,o}$ are the maximum and minimum of the truncated and observed components of the $j$th Gaussian variable, respectively.

\subsection{Full conditionals of latent positions and the tree scale}\label{ssec:samplelatent}
Let $\bT_{-j}$ be the submatrix of $\bT$ without the $j$th column. 
The full conditional distribution of  $\bt_{j}$ is given by,
\bfln{ 
    p(\bt_{j}|\bT_{-j},\bE,\sigsq) \propto  \left[ \prod_{k \neq j }  \{ \Phi( \bt_{k}^{\top} \bt_{j} )\}^{e_{kj}}  \{1- \Phi(\bt_{k}^{\top} \bt_{j} )\}^{1-e_{kj}} \right]  
    \N_{L}(\bt_{j}|\btheta_{j},\bPsi_{j}), \quad j=1,\ldots,p,
}
where $\btheta_{j}$ and $\bPsi_{j}$ are mean and covariance matrix of $\bt_{j}$ given $\bT_{-j}$ (detailed expressions are provided in Section \ref{supp:samplelatent} of the Supplementary Materials). We update $\bt_{j}$ using the data augmentation technique of \citet{albert1993bayesian} by introducing the auxiliary data $\by_{j} \in \R^{p-1}$. Let $\TN(\mu,\sigma^2,e)$ be  $\N(\mu,\sigma^2)$ truncated to be positive if $e=1$ and negative if $e=0$. Conditining on $\bT$, each component $y_{kj}$ of $\by_{j}$ follows $\TN(\bt_{k}^{\top}\bt_{j},1,e_{kj})$, and the resulting augmented pdf of $\bt_{j}$ and $\by_{j}$ is
\bfln{ 
p(\bt_{j},\by_{j}&|\bT_{-j},\bE,\sigsq) \propto \\
& \prod_{k \neq j}
\left\{ 1(y_{kj}>0, e_{kj}=1) + 1(y_{kj}<0,e_{kj}=0) \right\} \N(y_{kj}|\bt_{k}^{\top} \bt_{j}, 1) \N_{L}(\bt_{j}|\btheta_{j},\bPsi_{j}).
}
We obtain a posterior sample of $\bT$ by  alternately sampling $\by_{j}$ and $\bt_{j}$ for $j=1,\ldots,p$. Conditional on $\bT$, we independently sample $y_{kj}$ from $\TN(\bt_{k}^{\top} \bt_{j},1,e_{kj})$ for $k\neq j$. Then, conditional on $\by_{j}$, we have the Gaussian full conditional of $\bt_{j}$ as
\begin{flalign}\nonumber
p(\bt_{j} |\by_{j}, \bT_{-j},\bE,\sigsq) \propto 
\N_{p-1}(\by_{j}|\bT_{-j}^{\top} \bt_{j}, \bI_{p-1}) \N_{L}(\bt_{j}|\btheta_{j},\bPsi_{j}).
\end{flalign}
Accordingly, we draw $\bt_{j}$ from $\N_{L}( \bgamma_{j}, \bDelta_{j} )$, where
\bfln{
\bDelta_{j} = \left( \bT_{-j}\bT_{-j}^{\top} + \bPsi_{j}^{-1} \right)^{-1}, \quad
\bgamma_{j} = \bDelta_{j} 
\left( \bT_{-j} \by_{j} + \bPsi_{j}^{-1} \btheta_{j} \right).
}

The edge inclusion probabilities are updated as $\pi_{jk}=\Phi(\bt_{j}^{\top}\bt_{k})$, $1\leq j<k \leq p$.
Conditional on $\bT$, we sample the tree scale parameter $\sigsq$  from
\bfln{
\sigsq|\bT \sim \IG\left(pL/2 + a_{\sigsq}, \vect(\bT)^{\top}(\bH\otimes \bI_{L})^{-1}\vect(\bT)/2  + b_{\sigsq}\right),
}
where $\vect(\bT)$ is the vector obtained by stacking the columns of $\bT$ and $\otimes$ is the Kronecker product.

For sampling concentration matrix and graph, we follow the block Gibbs sampler of \citet{wang2015scaling}. For completeness, we provide the block Gibbs sampling algorithm in Section \ref{supp:spikeandslab} of the Supplementary Materials. Upon the completion of the MCMC, we compute the posterior mean of the edge inclusion probabilities for each pair of nodes,  $\widehat{\pi}_{jk}=\sum_{s=1}^{S} e_{jk}^{(s)}/S$, where the superscript indexes posterior samples. We obtain the estimated graph by selecting edges for which $\widehat{\pi}_{jk}$ is larger than some cutoff. 
We choose the cutoff to control the posterior expected FDR \citep{mitra2013bayesian,peterson2015bayesian} at prespecified level $\alpha$, where the posterior expected FDR is a decreasing function of cutoff $c$ defined as 
\bfl{\label{eq:postfdr}
\E(\text{FDR}_{c}|\text{data})=\frac{\sum_{j<k}(1-\widehat{\pi}_{jk})1(\widehat{\pi}_{jk}> c)}{\sum_{j<k}1(\widehat{\pi}_{jk}> c)}.
}


\section{Simulation}\label{sec:simulation}

We simulate microbiome data following the data generation mechanism proposed in \cite{yoon2019spring}, which allows to obtain synthetic samples that exactly follow the empirical marginal cumulative distributions of measured microbiome count data while respecting user-specified microbial dependencies via $\bOmega$.

Specifically, we randomly generate 10 phylogenetic trees $\Tcal_{1},\ldots,\Tcal_{10}$ with $p=50$ terminal nodes by the  \textsf{R} package \textsf{ape} \citep{paradis2019ape}, constituting 10 simulation scenarios. For each tree, the latent positions of terminal nodes, $\bt_{1},\ldots,\bt_{p}$, are generated from the diffusion process as described in Section \ref{ssec:tree} with $\sigma^2=3$ and $L=2$. The true graph adjacency matrix $\bE_0$ is obtained by independently generating $e_{jk} \sim \Bern(\pi_{jk})$ for $1\leq j<k \leq p$ with $\pi_{jk} = \Phi(\bt_{j}^{\top} \bt_{k})$. The trees and the true graphs are plotted in Section \ref{supp:phylotrees} of the Supplementary Materials.
Given $\bE_0$, the concentration matrix $\bOmega$ is drawn from G-Wishart$(\bI_{p},4)$ \citep{roverato2002hyper}. 
To obtain empirical cdfs, we use the quantitative microbiome profiling data of \citet{vandeputte2017quantitative} from $n=106$ subjects,  more detailed description of the data is provided in Section \ref{sec:application}. We select $p = 50$ genera (variables) of which 6 genera have no observed zero counts, and 44 genera have 20\% to 70\% zero counts across samples.

Given the empirical cdf $\widehat{F}_{j}$ of each selected genera, $j=1,\ldots, p$, and the concentration matrix $\bOmega$, we generate $n=106$ independent latent Gaussian vectors $\bz_{i}\sim\N_{p}(\zeros, \bOmega^{-1})$. The final data $\bx_1,\dots,\bx_n$ are obtained as
\bfln{
x_{ij} = \widehat{F}_{j}^{-1}\circ \Phi(z_{ij}), \quad i=1,\ldots,n, \quad j=1,\ldots,p.
}
We consider 50 independent replications of this data generating process for each scenario.

We compare the performance of the the proposed phylogenetically-informed Bayesian Copula Graphical model (PhyloBCG) with SPIEC-EASI \citep{kurtz2015sparse} and SPRING \citep{yoon2019spring}. Additionally, we also consider two special cases of PhyloBCG with the following simplification to the prior model of graph:
\begin{flalign*}
\moracle:~& \pi_{jk} = \binom{p}{2}^{-1}|\bE_0|;\\
\mdist:~&   \pi_{jk} = \exp(-\gamma d_{jk} ), \quad \gamma \sim \Exp(1); 
\end{flalign*}
where $|\bE_0|$ is the number of true edges in the underlying graph, and $d_{jk}$ is the tree distance between terminal nodes $j$ and $k$, defined as the sum of the branch lengths to their most recent common ancestor. We refer to the first model as ``$\moracle$'' since it uses the knowledge of true graph sparsity, however it does not use any tree information. We refer to the second model as ``$\mdist$'', as it directly incorporates tree distances between the terminal nodes: the edge inclusion probability is higher (smaller) when the corresponding nodes are closer (farther) to each other on the tree. $\mdist$ model thus takes into account the information from the tree, however, the tree information is deterministically incorporated to the model in contrast to the stochastic incorporation of PhyloBCG.

To implement SPIEC-EASI and SPRING, we use the corresponding R packages \citep{kurtz2017spieceasi,yoon2019springpackage} with sparsity parameters tuned over 100 values. 
For PhyloBCG, Oracle and Dist, the hyperparameters are fixed as $a_{\sigsq}=b_{\sigsq}=a_{v_{0}^{2}}=b_{v_{0}^{2}}=0.001$, $h=2500$, $\lambda=1$, and $L=2$. PhyloBCG is relatively robust to the hyperparameter setting; see the sensitivity analyses in Section \ref{supp:sensitivity} of the Supplementary Materials. We obtain an MCMC posterior sample of size $S=5000$ after 500 burn-in iterations. The graph estimate is obtained by thresholding the mean of posterior inclusion probability $\widehat{\pi}_{jk}$ at $c_{0.05}$, the smallest $c$ that controls the posterior expected FDR \eqref{eq:postfdr} at level 0.05. 

\begin{figure}[!t]
  \centering
  \includegraphics[width=1\textwidth]{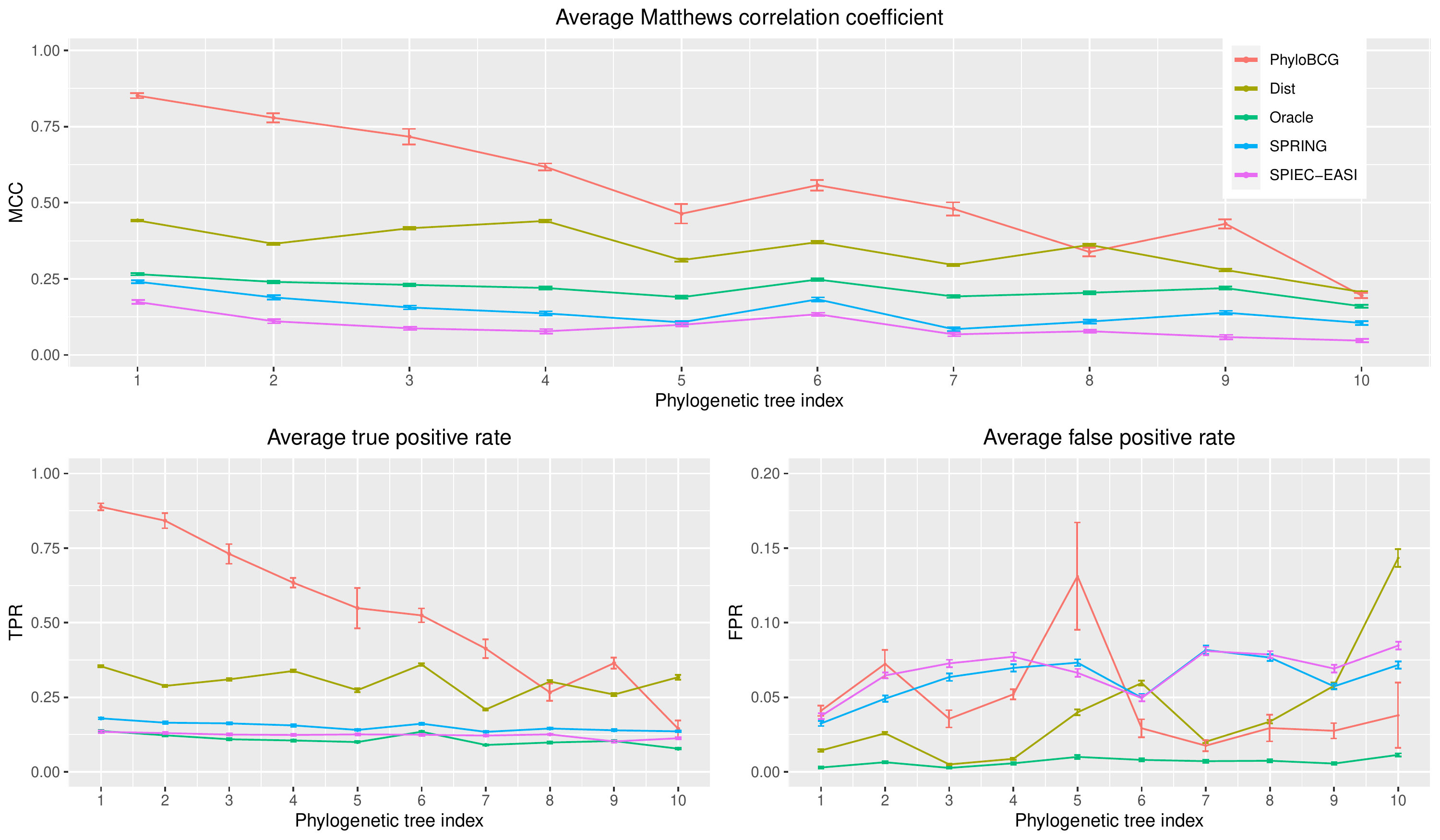}\\
  \caption{Averages of Matthews correlation coefficient (MCC), true positive rates (TPR) and false positive rates (FPR) with 2 standard error bars. The phylogenetic trees are ordered in term of the global clustering coefficients. Averages are taken over 50 replicated data sets.} \label{fig:simsum}
\end{figure}

We assess the graph recovery performance (accuracy in estimating $\bE_{0}$) of each method using  Matthews correlation coefficient \citep{matthews1975comparison}, true positive rate, and false positive rate, which will be denoted by MCC, TPR, and FPR, respectively. Ranging from -1 to 1, a larger value of MCC represents a better network estimation accuracy, where the two boundary values indicate completely correct (+1) and wrong (-1) edge selection, respectively.
Figure \ref{fig:simsum} summarizes the mean values of these metrics for each of the 10 phylogenetic trees based on 50 replications. The phylogenetic trees are sorted in terms of the global clustering coefficient \citep{wasserman1994social} of the true graph from largest value ($\Tcal_{1}$) to lowest value ($\Tcal_{10}$). A large value of the global clustering coefficient indicates a presence of microbial communities, with dense interactions within the same community and sparse interactions across communities. A small value of the global clustering coefficient indicates a random interaction pattern close to what will be expected with Erdos-Renyi random graph. Thus, we anticipate the phylogenetic tree to be more informative for network estimation when the global clustering coefficient is larger. 

Figure \ref{fig:simsum} supports that incorporating phylogenetic tree information improves the network estimation accuracy, with Phylo and Dist having higher MCC, higher TPR, and similar FPR values when compared to other methods. As expected, the value of MCC for the tree-based methods decreases as the phylogenetic tree becomes less informative (larger tree index). For PhyloBCG, this trend is driven by the decreasing TPR, whereas, for Dist, it is the increasing FPR. Although PhyloBCG and Dist both use the evolutionary information, PhyloBCG shows significantly better performances than Dist possibly due to the flexibility of the latent space embedding.


Note that the FPR of PhyloBCG has larger variability for trees $\Tcal_{5}$ and $\Tcal_{10}$ than others, with $\Tcal_5$ leading to the largest mean FPR. The reason for increased FPR in these settings is the discrepancy between the phylogenetic tree and the true graph. Recall that the true edge inclusion probabilities $\pi_{jk}$ are obtained from the Gaussian latent positions rather than directly from the tree,  thus allowing the true graph to deviate, sometimes significantly, from the phylogenetic tree. To illustrate this phenomenon, Figure \ref{fig:violin_hjk} in the Supplementary Materials shows the upper triangular part of the true tree correlation matrix $\bH$ defined in Section~\ref{ssec:tree} against the edge indicators for $\Tcal_{1}-\Tcal_{10}$. Both $\Tcal_{5}$ and $\Tcal_{10}$ show a large number of disconnected edges, $e_{jk}=0$, with high tree correlation values which may contribute to the increase in FPR. Nevertheless, for $\Tcal_{5}$, $\mphylo$ still shows favorable performance, having much larger TPRs than SPIEC-EASI and SPRING. Under $\Tcal_{10}$, PhyloBCG shows comparable performance to Oracle. Considering that the prior edge inclusion probability of $\moracle$ relies on the knowledge of the true number of connected edges, whereas the proposed PhyloBCG model does not and performs at least as good as the $\moracle$, we conclude that PhyloBCG adapts well to the unknown sparsity of the underlying graph.

\section{Application to quantitative gut microbiome profiling data}\label{sec:application}
\subsection{Data and phylogenetic tree}

\begin{figure}[t]
  \centering
  \includegraphics[width=1\textwidth]{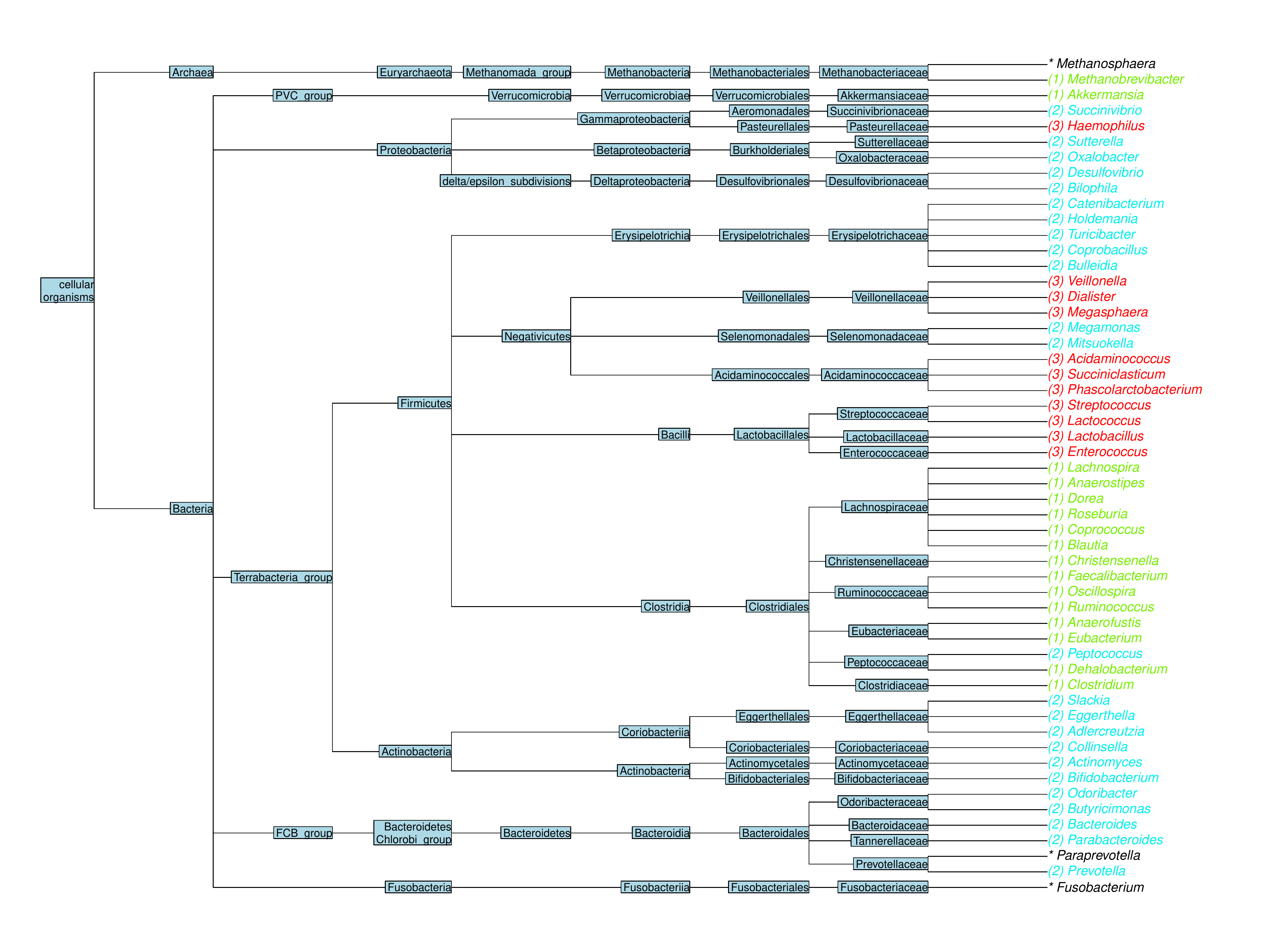}
    \caption{The phylogenetic tree of 54 genera. Different numbers (colors) represent three microbial communities derived from the graph estimated by the proposed PhyloBCG; the asterisk indicates a stand-alone node. }\label{fig:qmptreemembership}
\end{figure}

We focus on estimating genus-level association network of the QMP data \citep{vandeputte2017quantitative} that consists of $n=106$ healthy subjects' gut microbiome. We use the data as processed in \citet{yoon2019spring}, which can be obtained from the R-package \textsf{SPRING} \citep{yoon2019springpackage}. Among the 91 genera, there are 33 genera with missing names and 4 genera without available phylogenetic information on the National Center for Biotechnology Information (NCBI) database on which we based the phylogenetic tree construction. Consequently, we consider $p=54$ genera and obtain their phylogenetic tree based on the NCBI taxonomy database using the platform PhyloT\footnote{https://phylot.biobyte.de}. As the database does not provide divergence times of branches, we match the branch lengths with the taxonomic ranks as illustrated in Figure \ref{fig:qmptreemembership}. These genera have up to 70\% zeros (14 have more than 50\% zeros).

\subsection{Analysis}
For PhyloBCG, we use the same hyperparameter values as in Section \ref{sec:simulation} and run 4 parallel Markov chains for 100,000 iterations after 25,000 burn-in iterations. We then concatenate the 4 chains and obtain a posterior sample of size 10,000 by retaining every 40th iteration, from which we compute the posterior means $\widehat{\pi}_{jk}$ and $\widehat{\bOmega}$. 
The microbial association network is estimated by controlling the posterior expected FDR at 0.1 which results in $c_{0.1}=0.719$. 
SPIEC-EASI and SPRING are tuned using 100 sparsity parameter values. The default stability threshold \citep{liu2010stability} is changed from 0.1 to 0.2 to avoid overly sparse network estimates. The recovered networks are shown in Figure \ref{fig:qmpnetwork}. 

\afterpage{
\begin{figure}[ht!]
  \centering
  \includegraphics[width=0.91\textwidth]{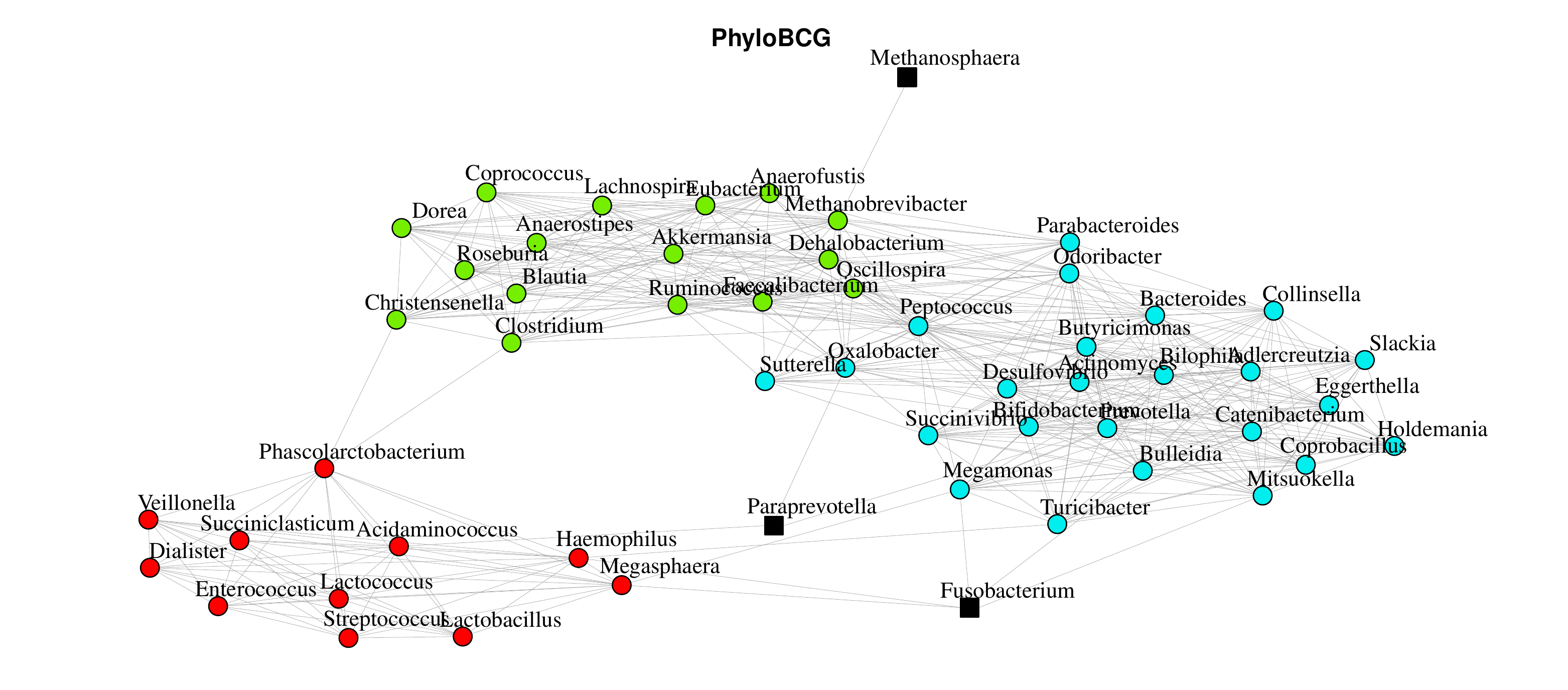}
  \includegraphics[width=0.90\textwidth]{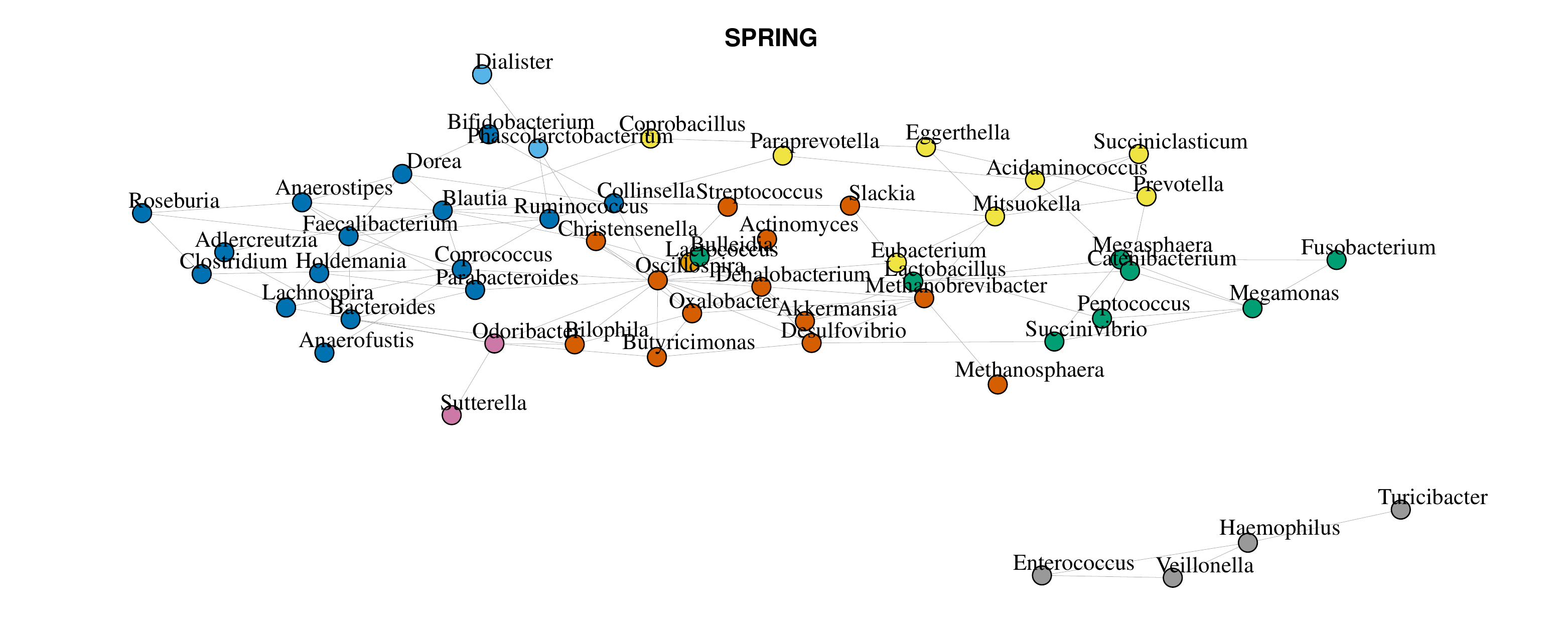}
  \includegraphics[width=0.90\textwidth]{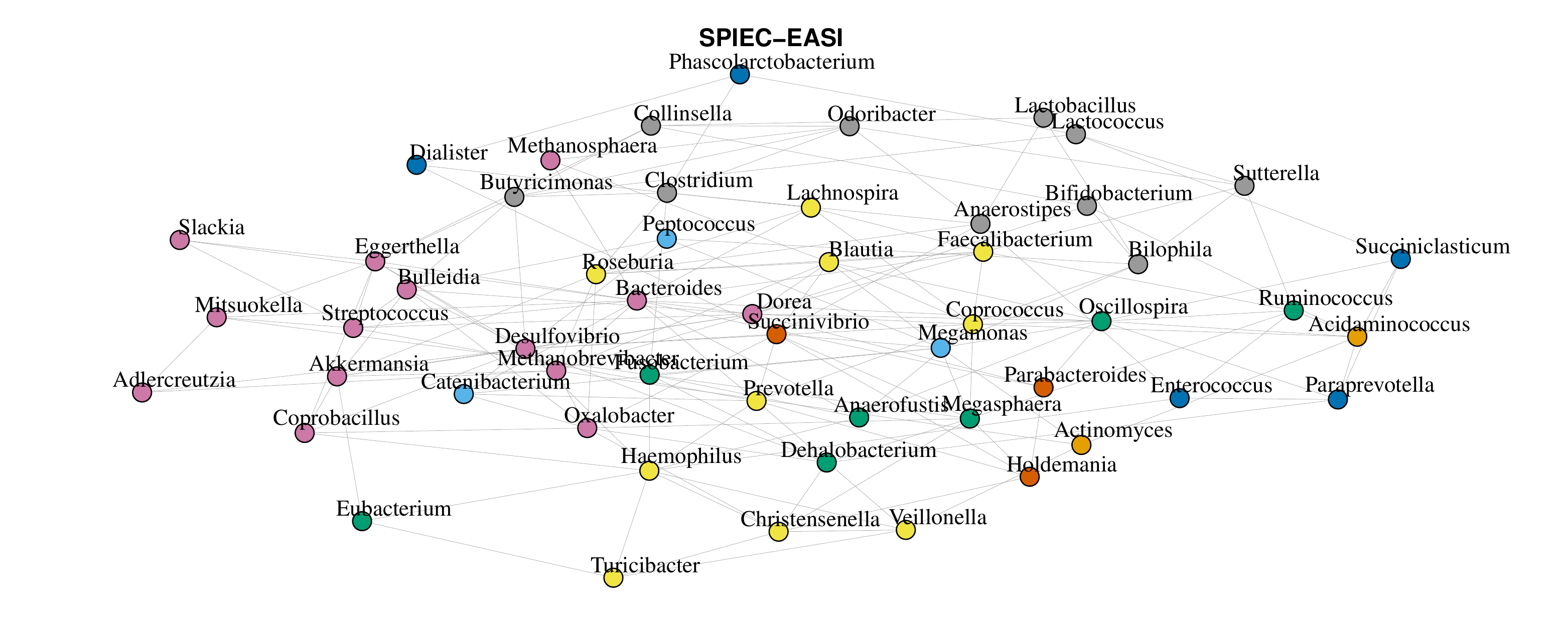}
    \caption{ Estimated microbial association networks of PhyloBCG, SPRING, and SPIEC-EASI with 54 genera in QMP data \citep{vandeputte2017quantitative}.
    Communities are separately estimated with the graph estimate of each model, and node colors represent the memberships.}\label{fig:qmpnetwork}
\end{figure}
}

\subsubsection{Overall network summary and interpretation}
We first compare estimated networks in terms of their density and community structure. The estimated network from PhyloBCG appears to be denser and have much more definitive communities than those from SPIEC-EASI and SPRING. 
 While we do not know the true network and community structure of the gut microbiota in the study population, the following reasons support our belief that the additional findings of microbial interactions and communities from PhyloBCG are biologically meaningful.

\begin{figure}[t!]
  \centering
  \includegraphics[width=1\textwidth]{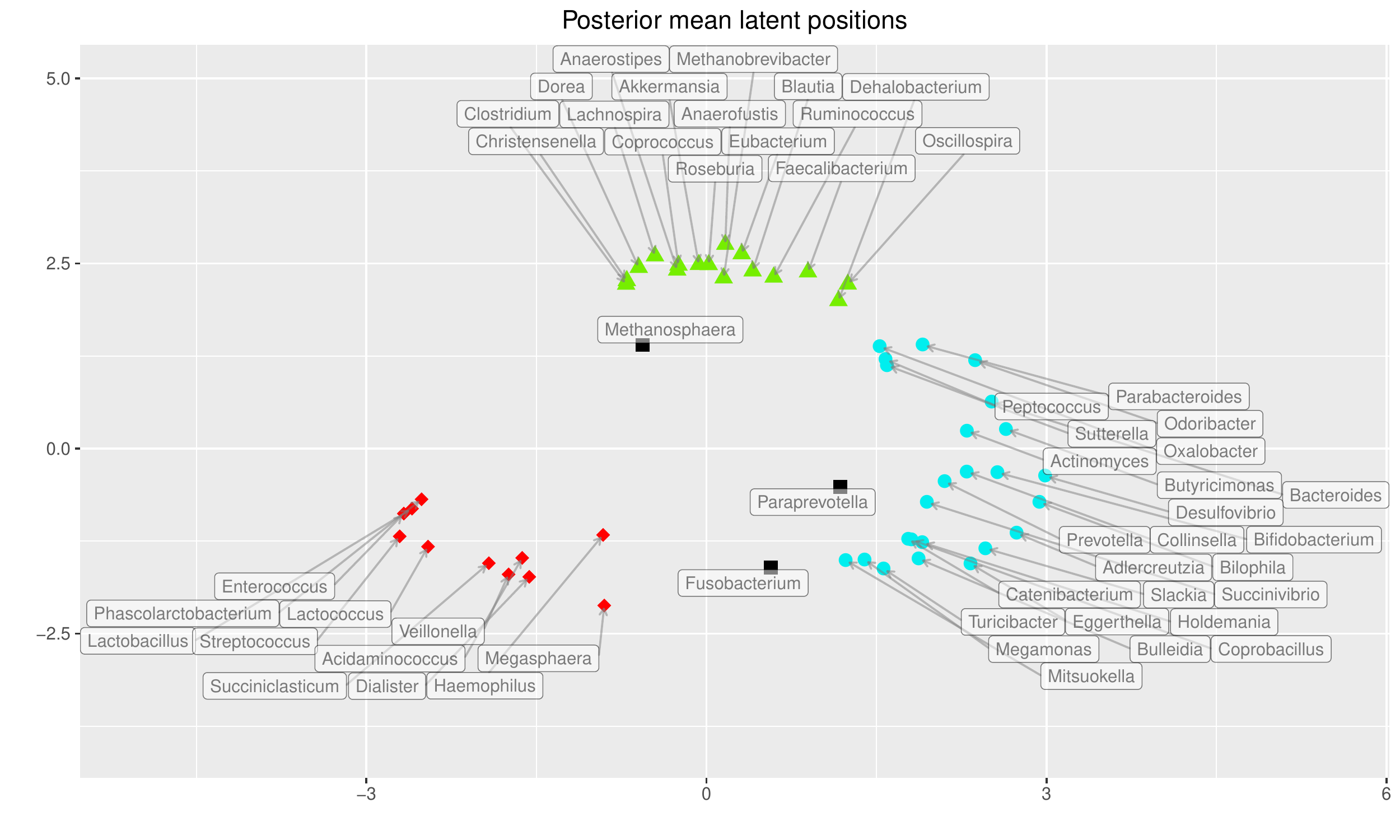}
    \caption{ Posterior mean latent positions of 54 genera in QMP data \citep{vandeputte2017quantitative}. Community memberships estimated with the graph estimate of PhyloBCG are marked by colored symbols. Black squares represent stand alone genera. }\label{fig:post_latentpositions}
\end{figure}

First, microbes are known to form communities \citep{pflughoeft2012human}. Applying the edge proximity measure of \citet{newman2004finding} to the estimated network by PhyloBCG, we find three evident microbial communities which are marked with different colors in Figures \ref{fig:qmptreemembership} and  \ref{fig:qmpnetwork}. Posterior mean latent positions, illustrated in Figure \ref{fig:post_latentpositions}, also form three distinct clusters that consistently match the estimated microbial communities.
Interestingly, the genera within each of these communities tend to share unique characteristics. On the one hand, most of the genera from the top two communities in Figure \ref{fig:qmptreemembership} are obligate anaerobes which only survive in the absence of oxygen. 
On the other hand, the community located at the bottom of Figure \ref{fig:qmptreemembership} contains genera with species that need or at least can tolerate oxygen. For example, \textit{Streptococcus}  and \textit{Enterococcus} contain facultative anaerobic species that are able to utilize oxygen as a source of energy, but can also generate energy anaerobically in an oxygen-deficient environment \citep{fisher2009ecology,clewell1981streptococcus}. All the members of \textit{Haemophilus} are facultatively anaerobic species or aerobic species, where aerobic species need oxygen to survive \citep{cooke20171611haemophilus}. \textit{Lactobacillus} contains aerotolerant and microaerophilic species \citep{zheng2020lactobacillus}. Aerotolerant species do not need oxygen and use anaerobic fermentation to generate energy, but oxygen is not toxic to them. Microaerophiles need oxygen to survive but require low oxygen concentration to thrive and are damaged by high oxygen level, e.g., atmospheric oxygen level. Furthermore, the bottom community has very few interactions with the two top communities, possibly because of their distinct living environments. By contrast, the estimated microbial interaction networks from SPIEC-EASI and SPRING do not present obvious communities; this lack of community structure is, based on existing literature \citep{rohr2014, peralta2016merging}, unlikely.  

Second, while the phylogenetic tree prior helps identify additional interactions and communities, it does not dictate the posterior inference. Some of the genera from the top right community in Figure \ref{fig:qmptreemembership} (e.g., \textit{Succinivibrio} and \textit{Prevotella}) are not phylogenetically similar to each other.  Similarly, the other two communities also contains phylogenetically distant genera. This suggests that the phylogenetic tree prior does not override the information of associations that are strongly supported by the data. 

Third, from the simulation study, we have seen that PhyloBCG is much more powerful in detecting interactions than SPIEC-EASI and SPRING especially when there is a clear community structure, while also having comparable FPR.
Given that the communities of the estimated network by PhyloBCG seem biologically plausible, the additional interactions found by PhyloBCG are more likely to be true positives than false discoveries, some of which will be explained in detail in the next section.

\subsubsection{Detailed explanation of interactions}
\textbf{All models} identify strong negative partial correlations between \textit{Dialister} and \textit{Phascolarctobacterium}. This finding is in agreement with multiple published results. The original QMP study \citep{vandeputte2017quantitative} finds a strong negative correlation between these two genera. \citet{naderpoor2019faecal} report that \textit{Phascolarctobacterium} (\textit{Dialister}) is positively (negatively) correlated with insulin sensitivity. Consistently, \citet{pedrogo2018gut} indirectly observe a trade-off relationship between the two genera from obese groups. 
Besides, strong positive partial correlations between \textit{Oscillospira} and \textit{Ruminococcus}  
are found by all methods as well, which is consistent with the finding in \citet{chen2020high}. 

\textbf{PhyloBCG and SPRING} detect relatively strong partial correlations for the pairs (\textit{Mitsuokella}, \textit{Prevotella}) and (\textit{Ruminococcus}, \textit{Blautia}). In the phylogenetic tree displayed in Figure \ref{fig:qmptreemembership}, \textit{Ruminococcus} and \textit{Blautia} are relatively close to each other, being the members of the same order, Clostridales. \textit{Mitsuokella} and \textit{Prevotella} are, however, phylogenetically distant from each other, indicating that the QMP data present a strong association between them and that the tree prior of the proposed PhyloBCG does not dominate the inference. These findings agree with the network analysis of \citet{ramayo2016phylogenetic}, where they also detect positive partial correlations in the pairs (\textit{Mitsuokella}, \textit{Prevotella}) and (\textit{Ruminococcus}, \textit{Blautia}). 

Additionally, the pairs (\textit{Oscillospira}, \textit{Butyricimonas}) and (\textit{Eubacterium}, \textit{Peptococcus}) also show relatively strong partial correlations under the two models. Both pairs are known to be related to  diet and leanness. \citet{oh2020association} discover positive correlations between body weight and each of \textit{Eubacterium} and \textit{Peptococcus}. \citet{garcia2018shifts} find negative correlations between unhealthy diet (high intake of saturated fat and refined carbohydrates) and each of \textit{Oscillospira} and \textit{Butyricimonas}.  
Furthermore, positive partial correlations for the pairs (\textit{Bacteroides}, \textit{Bilophila}) and (\textit{Akkermansia}, \textit{Methanobrevibacter}) are observed, where these results match the analyses in \citet{vandeputte2016stool} and \citet{vandeputte2017quantitative}, respectively.


\begin{table}[t]
\centering
\caption{Selected pairs of genera with strong association as identified by PhyloBCG. ``--'' indicates that no significant partial correlation is found by the corresponding.}\label{tab:pairs}
\resizebox{1\textwidth}{!}{
\renewcommand{\arraystretch}{1.3}
    \begin{tabular}{ll|ccc|l}
              \multicolumn{2}{c|}{\multirow{2}{*}{Pairs of Microbial Genera}}&\multicolumn{3}{c|}{Partial Correlations}& \multirow{2}{*}{Reference} \\\cline{3-5}
              && PhyloBCG & SPRING & SPIEC-EAIS & \\
              \hline
Dialister     & Phascolarctobacterium & -0.384   & -0.185 & -0.252 &  \citet{vandeputte2017quantitative}, \citet{naderpoor2019faecal}   \\
Oscillospira  & Ruminococcus       &  0.129   & 0.320  & 0.335  &  \citet{chen2020high}  \\
Mitsuokella   & Prevotella         &  0.155   & 0.152  & --     & \citet{ramayo2016phylogenetic} \\
Ruminococcus  & Blautia            &  0.115   & 0.048  & --     & \citet{ramayo2016phylogenetic} \\
Oscillospira  & Butyricimonas      &  0.155   & 0.335  & --     & \citet{garcia2018shifts}, \citet{thomaz2021influence} \\
Eubacterium   & Peptococcus        & 0.178    & 0.083  & --     & \citet{oh2020association}   \\
Bacteroides   & Bilophila          &  0.137   & 0.221  & --     & \citet{vandeputte2017quantitative}  \\
Akkermansia   & Methanobrevibacter &  0.202   & 0.354  & 0.054  & \citet{vandeputte2016stool}    \\
Blautia       & Methanobrevibacter & -0.062   & --     & -0.329 & \citet{garcia2018shifts}, \citet{muller2019distal}    \\
Prevotella    & Bacteroides        & -0.107   & --     & -0.019 & \cite{vandeputte2017quantitative} \\ 
              &                    &          &        &        & \citet{ley2016prevotella}, \citet{johnson2017microbiome}   \\
Veillonella   & Streptococcus      &  0.149   & --     & --     & \citet{anbalagan2017next}, \citet{chen2020gut}\\ 
              &                    &          &        &        &  \citet{den2013diversity}, \citet{egland2004interspecies} \\ 
              &                    &          &        &        &  \citet{zoetendal2012human}, \citet{van2014immunomodulatory}  \\
Bifidobacterium & Holdemania       &  -0.167  & --     & --     &  \citet{liu2017fructooligosaccharide}, \citet{yang2018dietary} \\
              &                    &          &        &        &  \citet{wang2020rational} \\
              \hline
    \end{tabular}
}
\end{table}

\textbf{PhyloBCG and SPIEC-EASI} find negative partial correlations for the pairs (\textit{Blautia}, \textit{Methanobrevibacter}) and (\textit{Prevotella}, \textit{Bacteroides}). \textit{Blautia} and \textit{Methanobrevibacter} are known to be positively and negatively related to dietary fiber intake, respectively \citep{garcia2018shifts}. \citet{muller2019distal} suggest that their inverse relationship is possibly due to substrate competition as both use hydrogen as energy source. 

For \textit{Prevotella} and \textit{Bacteroides}, \citet{lozupone2012diversity} find the trade-off between these two genera --  carbohydrates (including simple sugars) focused diet increases \textit{Prevotella} and decreases \textit{Bacteroides} whereas protein and fat focused diet has the opposite effects on them. Their trade-off relationship is also discussed in \citet{ley2016prevotella} and \citet{johnson2017microbiome}. On the contrary, \citet{vandeputte2017quantitative} claim that their negative association is an artifact of using compositional rather than quantitative microbiome data for analyses . 

\textbf{PhyloBCG} uniquely discovers positive partial correlations for the pairs (\textit{Veillonella}, \textit{Streptoccocus}) and (\textit{Bifidobacterium}, \textit{Holdemania}). The estimated positive partial correlation between \textit{Veillonella} and \textit{Streptoccocus} is consistent with that of the gut microbial network analysis of \citet{chen2020gut}. \citet{anbalagan2017next} demonstrate their mutualistic relationship: \textit{Streptoccocus} uses glucose as a source of carbon and release lactic acid, whereas \textit{Veillonella} utilizes lactic acid as carbon and energy source for growth. There are also many studies reporting their co-occurrence and mutualism \citep{den2013diversity,egland2004interspecies, zoetendal2012human,van2014immunomodulatory}. 
For \textit{Bifidobacterium} and \textit{Holdemania}, \citet{liu2017fructooligosaccharide} report that prebiotic supplement significantly increases relative abundance of beneficial \textit{Bifidobacterium} and decreases \textit{Holdemania}, where \textit{Holdemania} is reported to be associated with unhealthy gut and antibiotic use \citep{yang2018dietary}. \citet{wang2020rational} discuss the underlying mechanism of the trade-off relationship. 
In summary, we find these uniquely identified pairs by the proposed PhyloBCG  to be well supported by existing literature.
All the genera pairs discussed above are summarized in Table \ref{tab:pairs} with their estimated partial correlations and supporting references.

\section{Discussion}
In this work, we propose a phylogenetically informed Bayesian truncated copula graphical model for estimating microbial association networks with QMP data. The proposed method explicitly accounts for the zero-inflated nature of the QMP data and incorporates the microbial evolutionary information through the diffusion process and latent position model.

Simulation study with various phylogenetic tree structures reveals that the phylogenetic prior significantly improves network estimation accuracy. In particular, the proposed model shows much larger true positive rates while having comparable false positive rates to existing microbial network estimation models. Our sensitivity analysis shows reasonably stable performance under various hyperparameter settings. Also, we find that the proposed model is robust to tree misspecification in that the phylogenetic tree does not override conditional dependence supported by data. In application to the QMP data analysis, the proposed model identifies several unique genus-level conditional dependencies that are missed by existing microbial network estimation methods. 

Although the proposed model is developed for microbial association network estimation, the established formulation of the truncated Gaussian copula can be directly applied to other zero-inflated data, such as single-cell  RNA sequencing data. Furthermore, the framework for incorporating evolutionary information is not limited to undirected graph estimation and can be extended to the directed graph estimation models. 

\section*{Acknowledgements}
The authors thank Grace Yoon for constructive discussions on QMP data analysis.
This work has been partially supported by the Texas A\&M Institute of Data Science (TAMIDS) and the Texas A\&M Strategic Transformative Research Program. Gaynanova's research was partially supported by the National Science Foundation (NSF CAREER DMS-2044823). Ni's research was partially supported by the National Science Foundation (NSF DMS-1918851). The \textsf{R} programs that reproduce the presented results are available at \url{https://github.com/heech31/phyloBCG}.

\appendix
\section{Gibbs sampling steps}\label{supp:gibbssteps}
\subsection{Sampling truncated observations and thresholds}\label{supp:samplezt}
Without loss of generality, let $\bz_{i}^{\top}=(\bz_{i,t}^{\top},\bz_{i,o}^{\top})$, where $\bz_{i,t}\in\R^{p_{i,t}}$, $\bz_{i,o}\in\R^{p_{i,o}}$, $p_{i,t}+p_{i,o}=p$, and $i=1,\ldots,n$. Also, let $\bSigma=\bOmega^{-1}$ be the covariance matrix of $\bz$. Conditioning on $\bz_{i,o}=\widehat{\bz}_{i,o}$, $\bz_{i,t}$ follows a truncated multivariate Gaussian distribution with pdf
\bfl{\label{eq:zit_conditionaldensity}
	p(\bz_{i,t}|\widehat{\bz}_{i,o},\bz_{i,t}<\bdelta_{t},\bOmega)= 
	\frac{\N_{p_{i,t}}(\bz_{i,t}|\bmu_{i},\bDelta_{i}) }{\PP(\bz_{i,t}<\bdelta_{t}|\widehat{\bz}_{i,o},\bOmega,\bdelta)}  1(\bz_{i,t}<\bdelta_{i,t}),
}
where 
\bfln{
\bmu_{i}=\bSigma_{i,ot}^{\top} \bSigma_{i,o}^{-1}\widehat{\bz}_{i,o}, \quad \bDelta_{i}=\bSigma_{i,t} - \bSigma_{i,ot}^{\top} \bSigma_{i,o}^{-1}
\bSigma_{i,ot},
}
and $\var(\bz_{i,t}) = \bSigma_{i,t}$, $\var(\bz_{i,o}) = \bSigma_{i,o}$ and $\cov(\bz_{i,o},\bz_{i,t}) = \bSigma_{i,ot}$.
We update $\bz_{i,t}$ by sampling from \eqref{eq:zit_conditionaldensity} for $i=1,\ldots,n$.

Let $\textup{Unif}(a,b)$ be the uniform distribution on the interval $(a,b)$. Also, let ${z}_{j,t}^{\max}$ $(\widehat{z}_{j,o}^{\min})$ be the maximum (minimum) of the truncated (observed) components of the $j$th Gaussian variable such that ${z}_{j,t}^{\max}=\max_{i} {z}_{ij,t}$ and $\widehat{z}_{j,o}^{\min}=\min_{i} \widehat{z}_{ij,o}$. We sample $\bdelta=(\delta_{1},\ldots,\delta_{p})^{\top}$ from $\delta_{j} \overset{ind}{\sim}\textup{Unif}({z}_{j,t}^{\max}, \widehat{z}_{j,o}^{\min})$, $j=1,\ldots,p$.

\subsection{Sampling latent positions and the tree scale}\label{supp:samplelatent}
Recall $\bT=[\bt_{1},\ldots,\bt_{p}]$ be the column-wise concatenation of the $p$ latent positions. We can write $\vect(\bT) \sim \N_{pL}(\zeros,\bPsi)$, where $\bPsi=\sigsq\bH\otimes \bI_{L}$.  
Let $\bT_{-j}$ be the submatrix of $\bT$ without the $j$th column. Then, $\bt_{j}|\bT_{-j}$ follows $\N_{L}(\btheta_{j},\bPsi_{j})$, where
\bfln{
\btheta_{j}=\bPsi_{j,-j} \bPsi_{-j,-j}^{-1} \vect(\bT_{-j}), \quad \bPsi_{j} = \bPsi_{j,j} -  \bPsi_{j,-j}\bPsi_{-j,-j}^{-1} \bPsi_{j,-j}^{\top},
}
$\var(\bt_{j})=\bPsi_{j,j}$, $\var\{\vect(\bT_{-j})\}=\bPsi_{-j,-j}$, and $\cov\{\bt_{j},\vect(\bT_{-j}) \}=\bPsi_{j,-j}$. The augmented pdf of $\bt_{j}$ and $\by_{j}$ is
\bfln{ 
p(\bt_{j},\by_{j}&|\bT_{-j},\bE,\sigsq) \propto \\
& \prod_{k \neq j}
\left\{ 1(y_{kj}>0, e_{kj}=1) + 1(y_{kj}<0,e_{kj}=0) \right\} \N(y_{kj}|\bt_{k}^{\top} \bt_{j}, 1) \N_{L}(\bt_{j}|\btheta_{j},\bPsi_{j}).
}

For $j=1,\ldots,p$, we alternate sampling $\by_{j}$ and $\bt_{j}$ as $y_{kj} \overset{ind}{\sim}\TN(\bt_{k}^{\top} \bt_{j},1,e_{kj})$ for $k\neq j$, and $\bt_{j} \sim \N_{L}( \bgamma_{j}, \bDelta_{j} )$, where
\bfln{
\bDelta_{j} = \left( \bT_{-j}\bT_{-j}^{\top} + \bPsi_{j}^{-1} \right)^{-1}, \quad
\bgamma_{j} = \bDelta_{j} 
\left( \bT_{-j} \by_{j} + \bPsi_{j}^{-1} \btheta_{j} \right).
}
We then update edge inclusion probabilities as $\pi_{jk}=\Phi(\bt_{j}^{\top}\bt_{k})$, $1\leq j<k \leq p$.
The tree scale parameter $\sigsq$  is drawn from
$\sigsq|\bT \sim \IG\left(pL/2 + a_{\sigsq}, \vect(\bT)^{\top}(\bH\otimes \bI_{L})^{-1}\vect(\bT)/2  + b_{\sigsq}\right)$.

\subsection{Sampling concentration matrix and graph}\label{supp:spikeandslab}
We follow the block Gibbs sampler of \citet{wang2015scaling}. For completeness, we provide the Gibbs sampling algorithm as follows. Recall that $\bOmega$ is the concentration matrix and $\bZ=[\bz_{1},\ldots,\bz_{p}]^{\top}$. Let $\bS=\bZ^{\top}\bZ$ and $\bV = [v_{jk}^2]_{1\leq j,k \leq p}$ be a zero-diagonal symmetric matrix with off-diagonal components $v_{jk}^2=v_{0}^2$ if $\omega_{jk}=0$ and $v_{jk}^2=hv_{0}^2$, otherwise. We consider the following partitions of $\bS$, $\bV$, and $\bOmega$:
\bfln{
\bS =
\begin{pmatrix}
\bS_{11} & \bs_{12}\\
\bs_{12}^{\top} & s_{22}\\
\end{pmatrix},
\quad
\bV =
\begin{pmatrix}
\bV_{11} & \bv_{12}\\
\bv_{12}^{\top} & 0\\
\end{pmatrix},
\quad
\bOmega =
\begin{pmatrix}
\bOmega_{11} & \bomega_{12}\\
\bomega_{12}^{\top} & \omega_{22}\\
\end{pmatrix}.
}
Let $\bu = \bomega_{12}$ and $v=\omega_{22} - \bomega_{12}^{\top} \bOmega_{11}^{-1} \bomega_{12}$. Then, by Proposition 1 of \citet{wang2015scaling}, we have the full conditional distributions as
\bfl{\label{suppeq:SSSL}
\bu|\bZ,\bOmega_{11},\bE,\bV \sim \N_{p-1}(-\bC\bs_{12},\bC), \quad v|\bZ \sim \textup{Gamma}(\frac{1}{2} +1, \frac{s_{22}+\lambda}{2}),
}
where $\bC=\{(s_{22}+\lambda)\bOmega^{-1} + \diag(\bv_{12}^{-1})\}^{-1}$ and the components of $\bv_{12}^{-1}$ are reciprocals of that of $\bv_{12}$. We update $\bOmega$ by sampling each column from \eqref{suppeq:SSSL}. The full conditional distribution of $e_{jk}$ is a Bernoulli with
\bfl{\label{eq:fullcondi_ejk_prob}
\Pr(e_{jk}|\bZ,\bOmega,v_{0}^2) =
\frac{\N(\omega_{jk}|0,hv_{0}^{2})\pi_{jk}}{\N(\omega_{jk}|0,hv_{0}^{2})\pi_{jk} + \N(\omega_{jk}|0,v_{0}^{2})(1-\pi_{jk})}, \quad 1\leq j<k\leq p.
}
Updating $e_{jk}$, $1\leq j< k \leq p$, by drawing from the corresponding Bernoulli distributions completes the Gibbs sampling steps for $\bOmega$ and $\bE$. We sample $v_{0}^2$  from
\bfln{
v_{0}^2|\bOmega,\bE \sim \IG\left(  \frac{p(p-1)}{4} + a_{v_{0}^2},\sum_{j<k} \frac{\omega_{jk}^{2}}{2h^{e_{jk}}}  + b_{\sigsq}\right).
}

\section{Sensitivity analysis}\label{supp:sensitivity}
\begin{figure}[h!]
	\centering
	\includegraphics[width=1\textwidth]{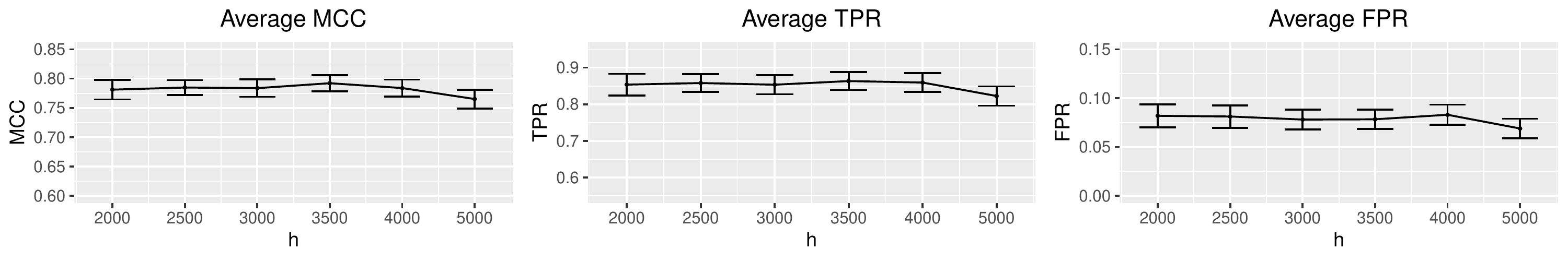}
	\includegraphics[width=1\textwidth]{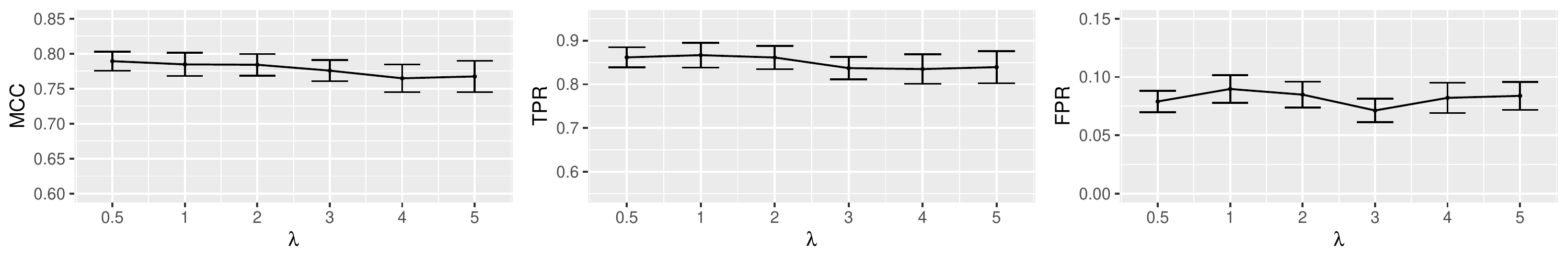}
	\includegraphics[width=1\textwidth]{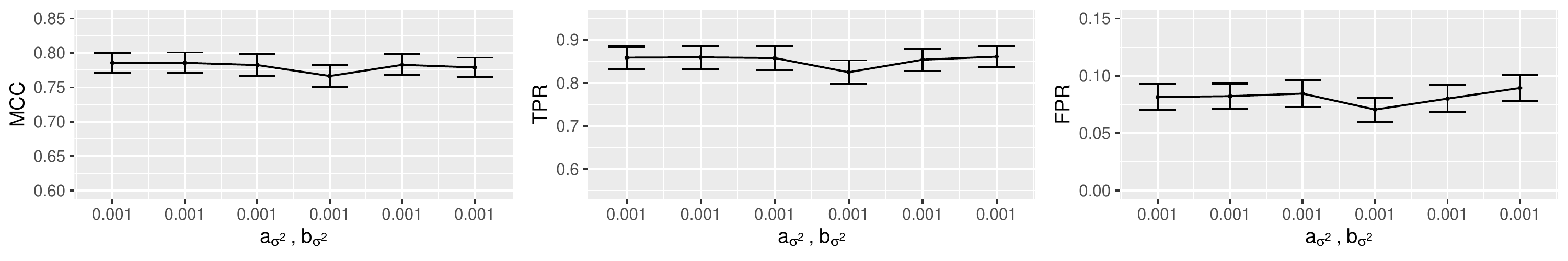}
	\includegraphics[width=1\textwidth]{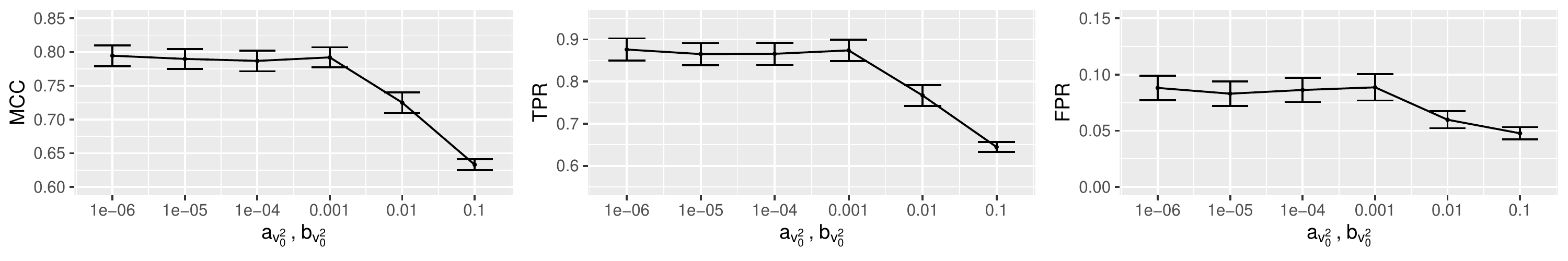}
	\caption{Summary of sensitivity analysis results with $\Tcal_{2}$. Average MCC, TPR, FPR, and FDR with 2 standard error bars based on 50 replicated data sets are illustrated. }\label{fig:sensitivity}
\end{figure}

With $\Tcal_{2}$ and corresponding $\bOmega$ and $\bE$, we examine the sensitivity of PhyloBCG to the changes in hyperparameters, $(a_{\sigsq},b_{\sigsq})$, $(a_{v_{0}^{2}},b_{v_{0}^{2}})$, $h$, and $\lambda$. We consider 6 different values for each hyperparameter while fixing the other hyperparameters 
at the default setting, $a_{\sigsq}=b_{\sigsq}=a_{v_{0^{2}}}=b_{v_{0}^{2}}=0.001$, $h=2500$, and $\lambda=1$. Specifically, we consider
\bfln{
h&=2000, 2500, 3000, 3500, 4000, 5000,\\
\lambda &= 0.5,1,2,3,4,5, \\
a_{\sigsq}&=b_{\sigsq} = 10^{-d}, \quad d=1,\ldots,6,\\
a_{v_{0^{2}}}&=b_{v_{0}^{2}} = 10^{-d}, \quad d=1,\ldots,6,
}
which result in 24 settings of hyperparameters. For each setting, we obtain a posterior sample of size 5000 after 500 burn-in iterations. Estimated graphs are obtained as specified in Section \ref{sec:simulation} based on 50 replicated data sets.

Figure \ref{fig:sensitivity} summarizes average Matthews correlation coefficient \citep{matthews1975comparison}, true positive rate, false-positive rate, and false-discovery rate denoted by MCC, TPR, and FPR, respectively. Overall, the results are very stable under the different choices of hyperparameters. Except for the two settings of $(a_{v_{0}^{2}},b_{v_{0}^{2}})$, PhyloBCG produces practically the same results across all settings of hyperparameters. For the two settings, $a_{v_{0}^{2}}=b_{v_{0}^{2}}=0.01, 0.1$, we observe that the average MCC values  decrease due to the decrease in TPR. This is because, larger values of $a_{v_{0}^{2}}=b_{v_{0}^{2}}$ make the inverse gamma prior put more mass on a large value of $v_{0}^2$. Thus, a large value of sampled $v_{0}^2$ decreases the success probabilities of edge indicators given in \eqref{eq:fullcondi_ejk_prob}. Therefore, we recommend to use $a_{v_{0}^{2}}=b_{v_{0}^{2}} \leq 0.001$.

\newpage

\section{Figures}\label{supp:phylotrees}
~

\begin{figure}[!ht]
	\centering
	\includegraphics[width=1\textwidth]{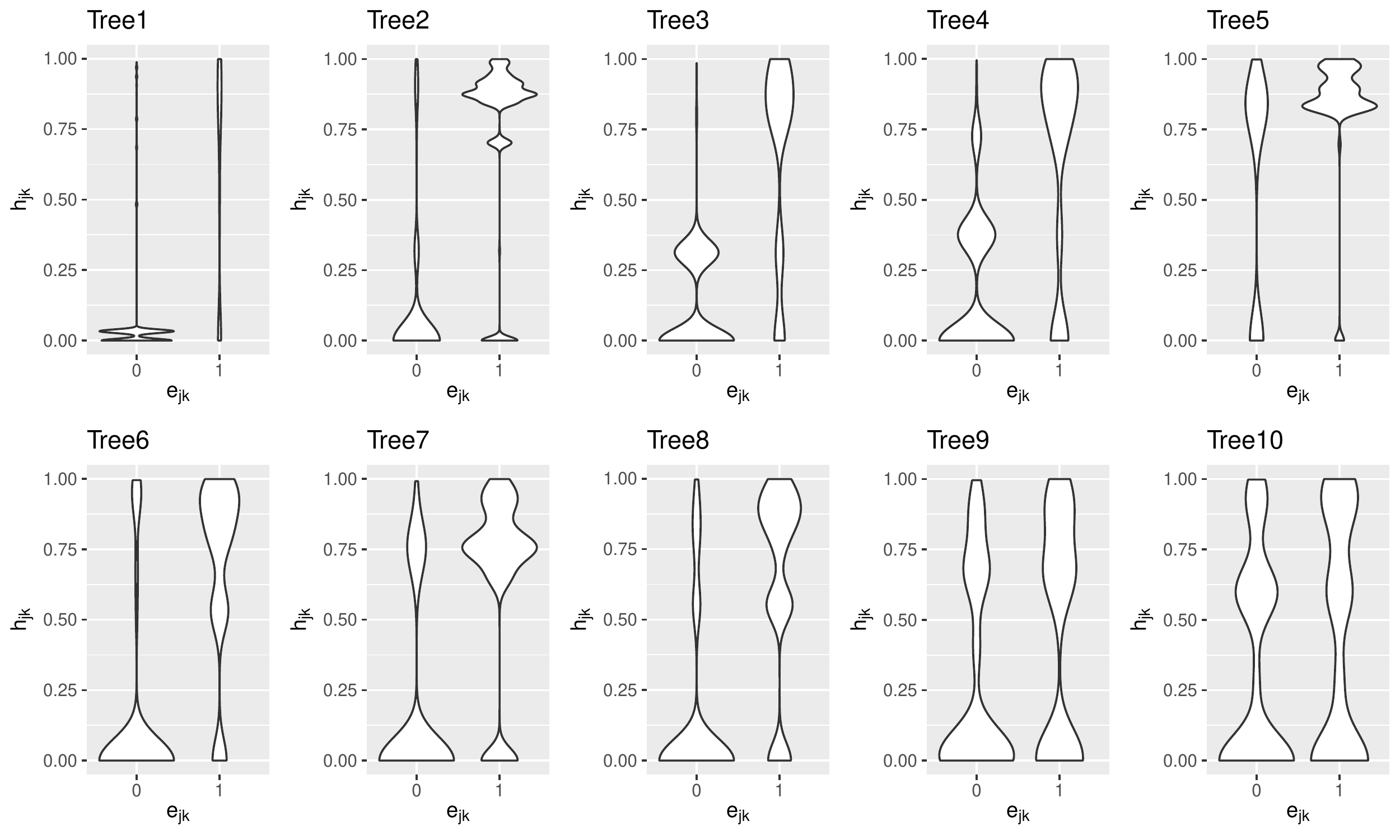}
	\caption{Violin plot of strict upper triangular components of the tree covariance matrix $\bH$ against the simulation true edge indicators. Two highly correlated terminal nodes, large $h_{jk}$, are desirable to be connected , $e_{jk}=1$.   }\label{fig:violin_hjk}
\end{figure}

\begin{figure}[ht]
	\centering
	\includegraphics[width=1\textwidth]{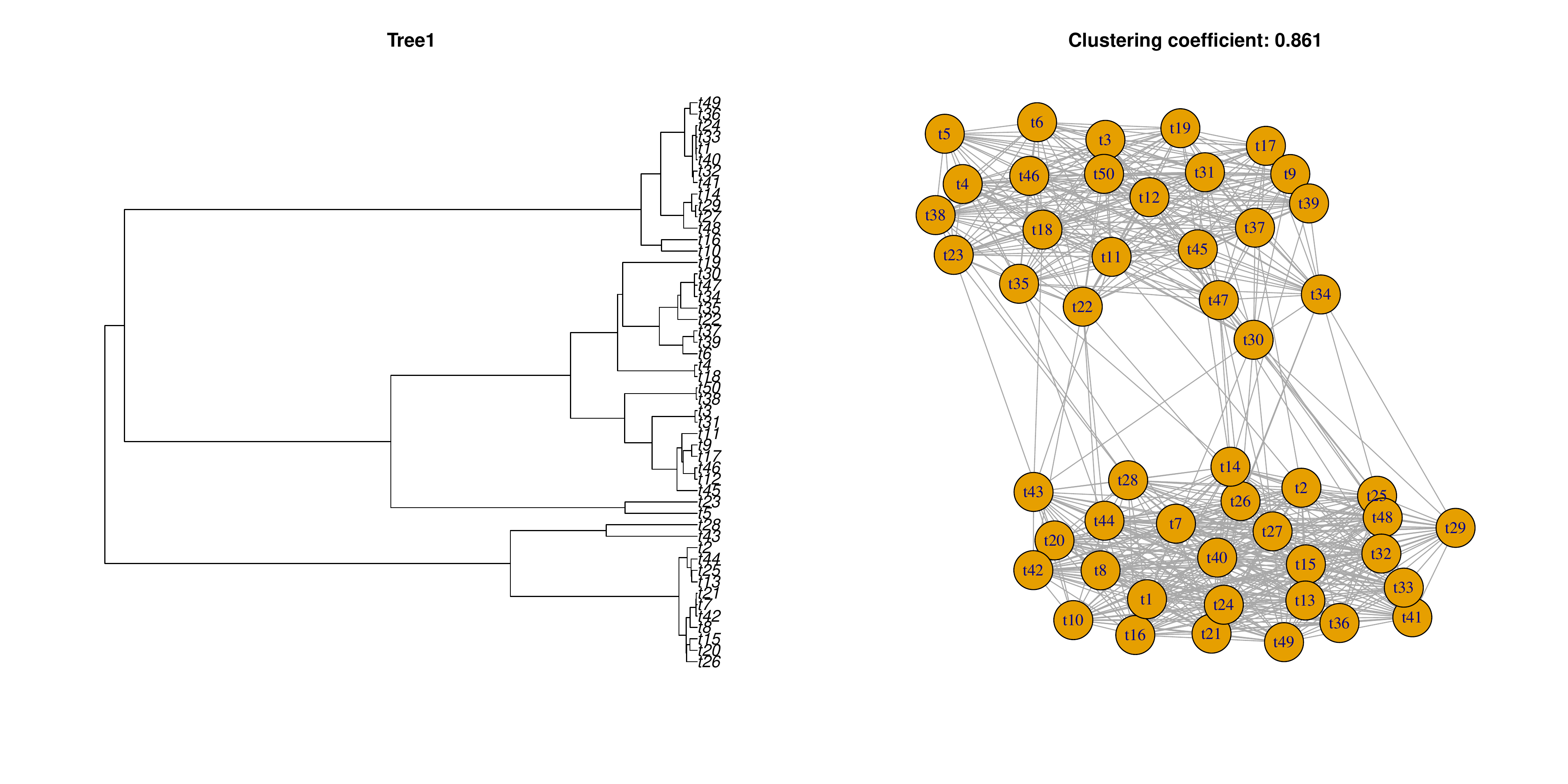}
	\includegraphics[width=1\textwidth]{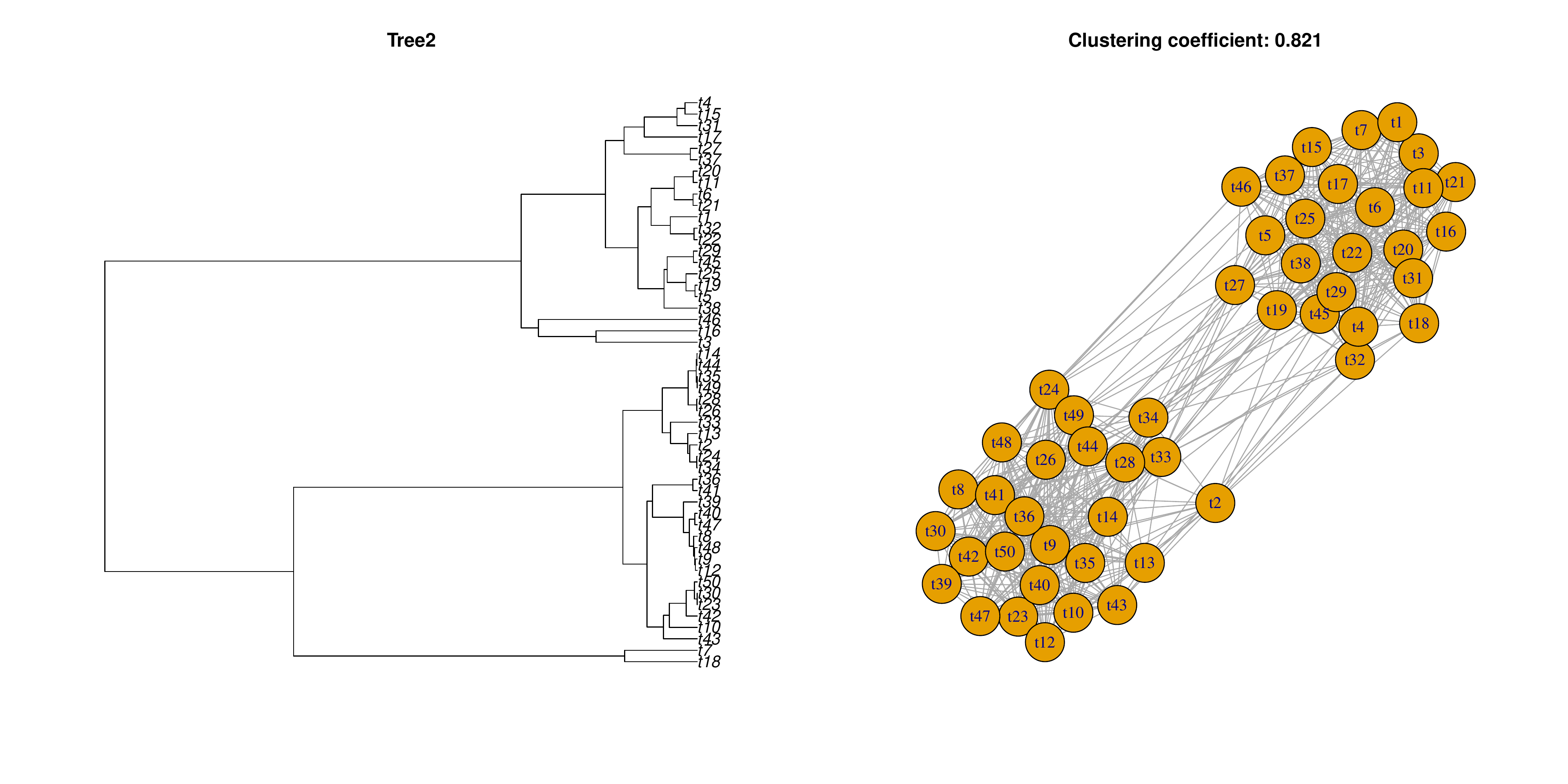}
	\caption{ $\Tcal_{1}$ and $\Tcal_{2}$}\label{fig:graph12}
\end{figure}

\begin{figure}[ht]
	\centering
	\includegraphics[width=1\textwidth]{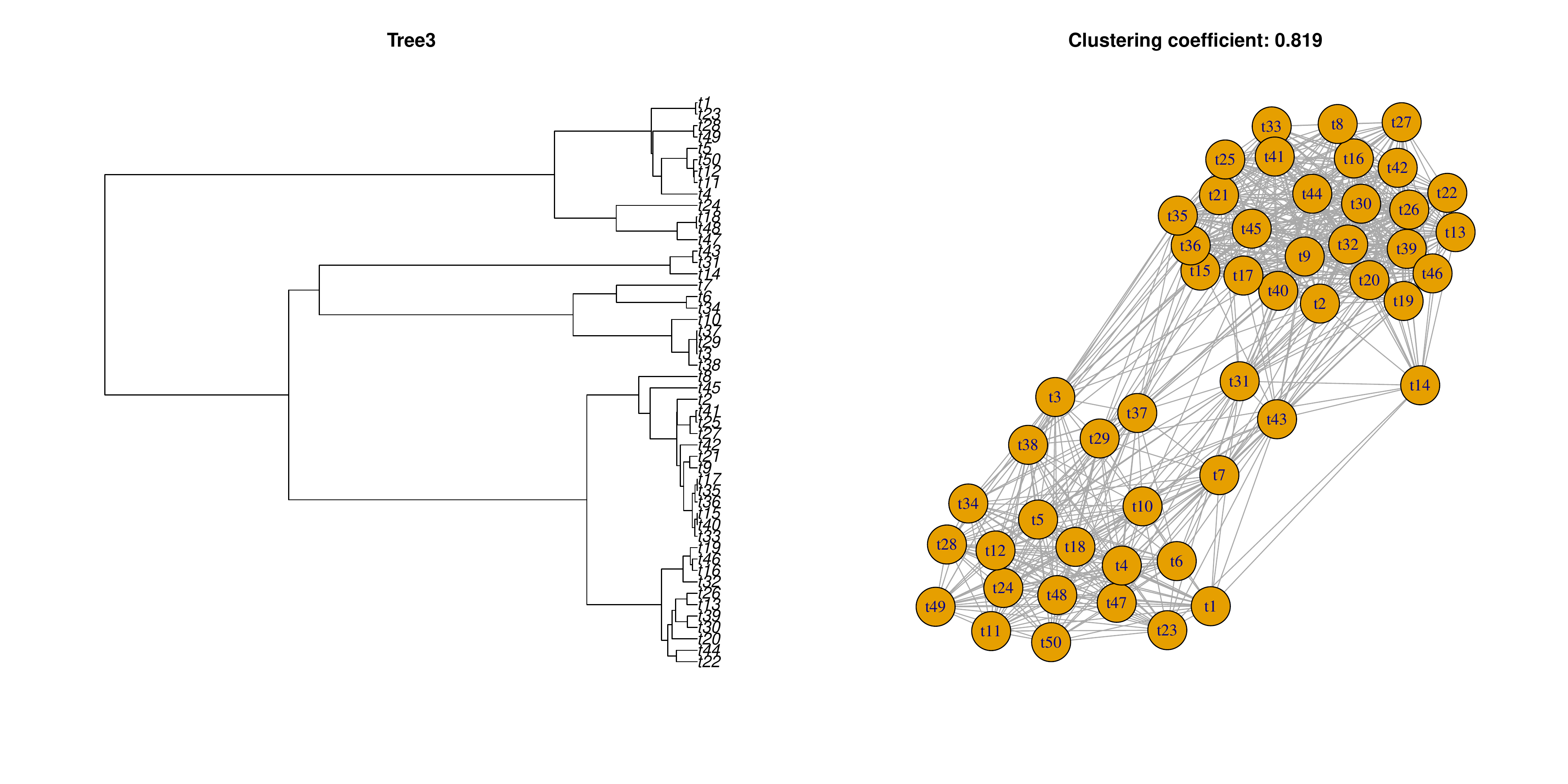}
	\includegraphics[width=1\textwidth]{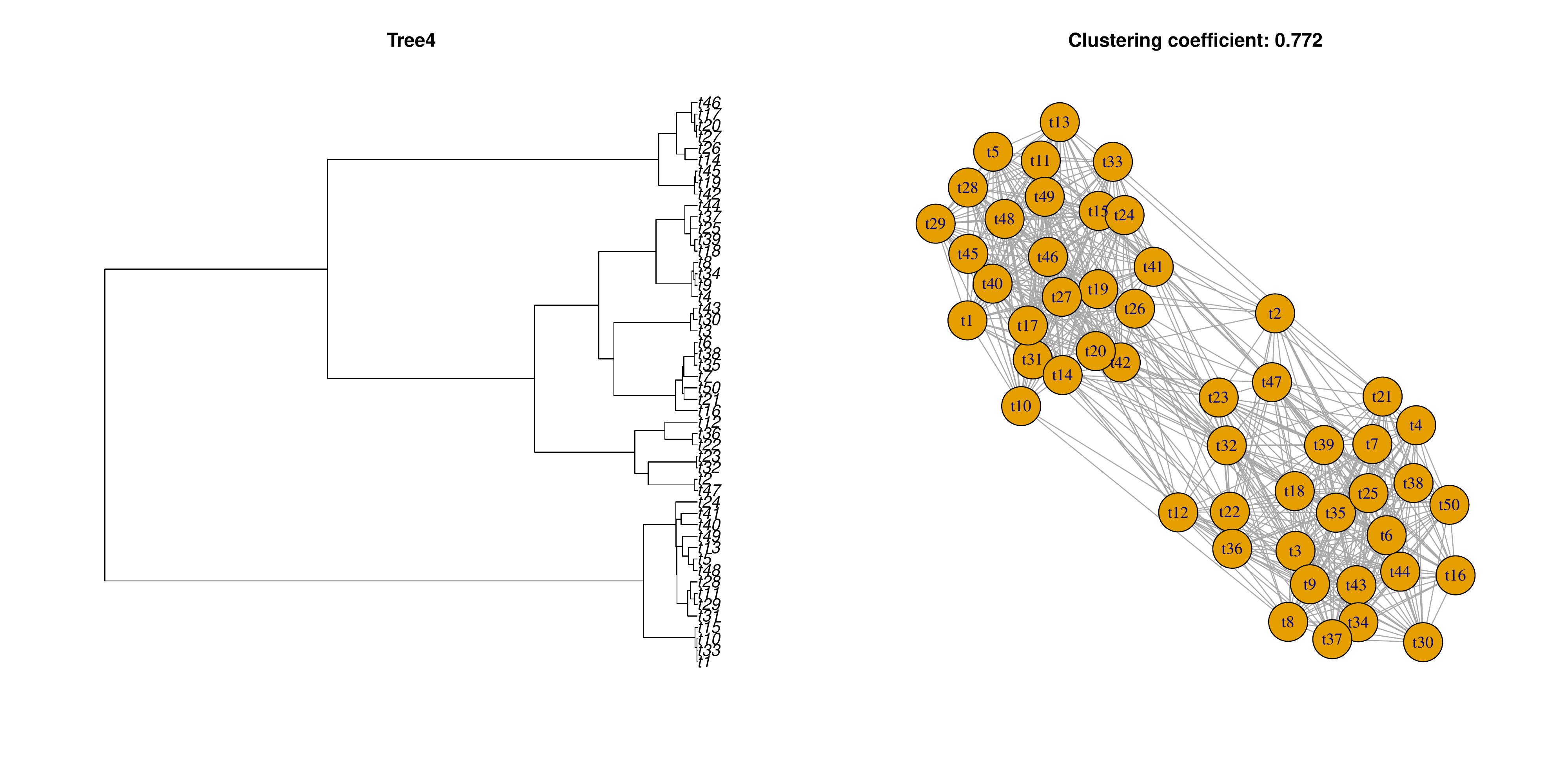}
	\caption{ $\Tcal_{3}$ and $\Tcal_{4}$ }\label{fig:graph34}
\end{figure}

\begin{figure}[ht]
	\centering
	\includegraphics[width=1\textwidth]{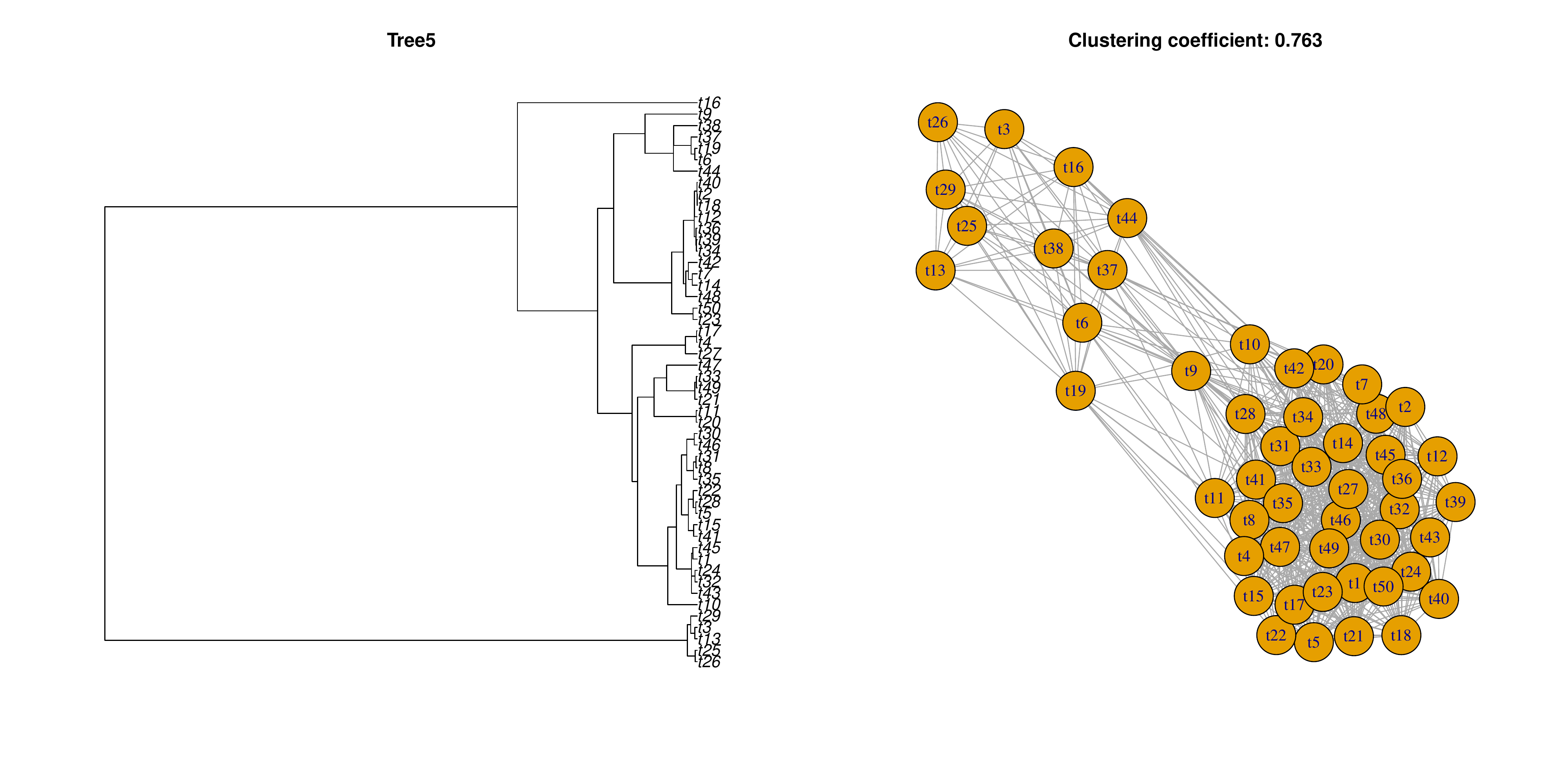}
	\includegraphics[width=1\textwidth]{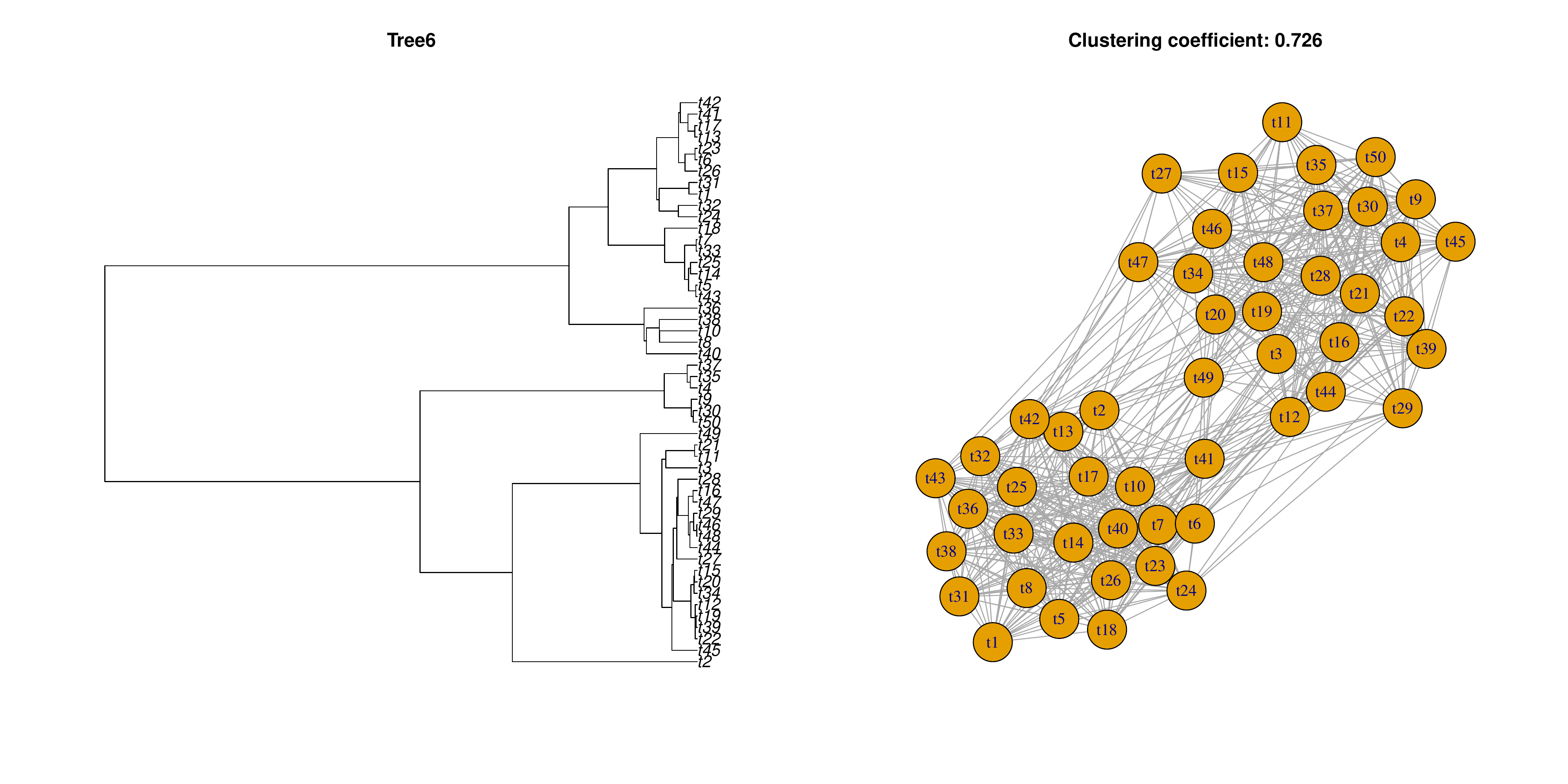}
	\caption{ $\Tcal_{5}$ and $\Tcal_{6}$ }\label{fig:graph56}
\end{figure}

\begin{figure}[ht]
	\centering
	\includegraphics[width=1\textwidth]{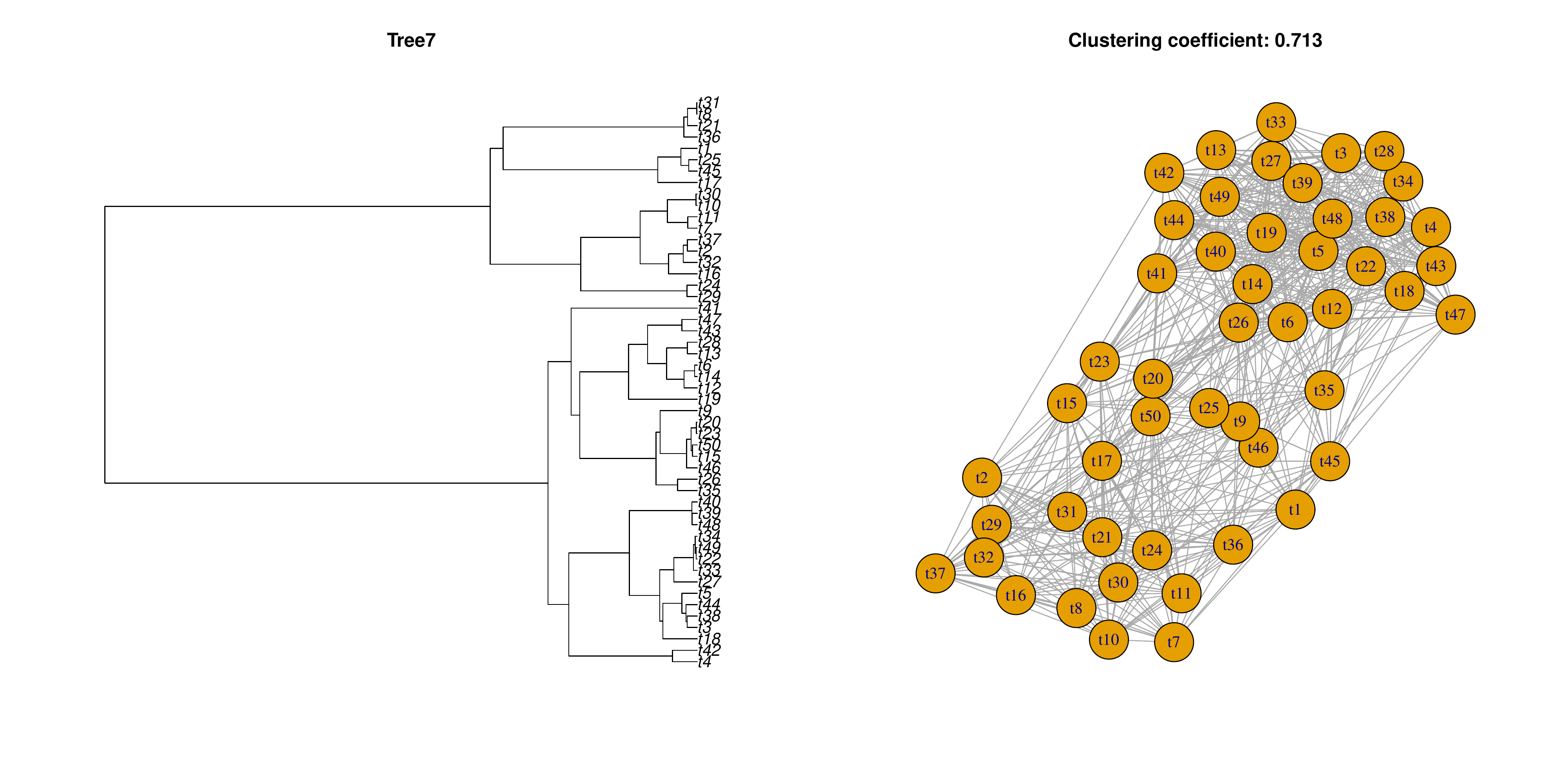}
	\includegraphics[width=1\textwidth]{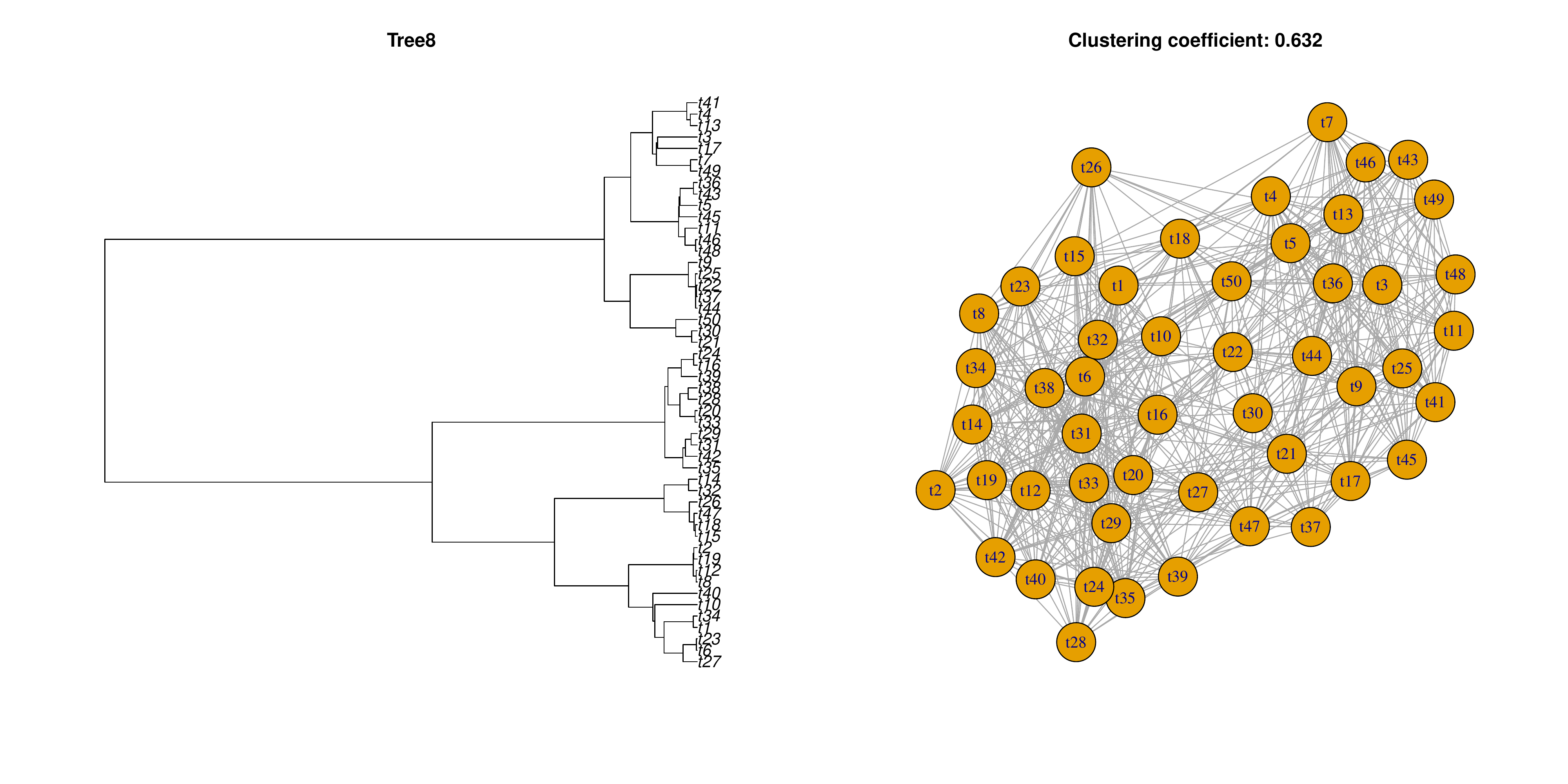}
	\caption{ $\Tcal_{7}$ and $\Tcal_{8}$ }\label{fig:graph78}
\end{figure}

\begin{figure}[ht]
	\centering
	\includegraphics[width=1\textwidth]{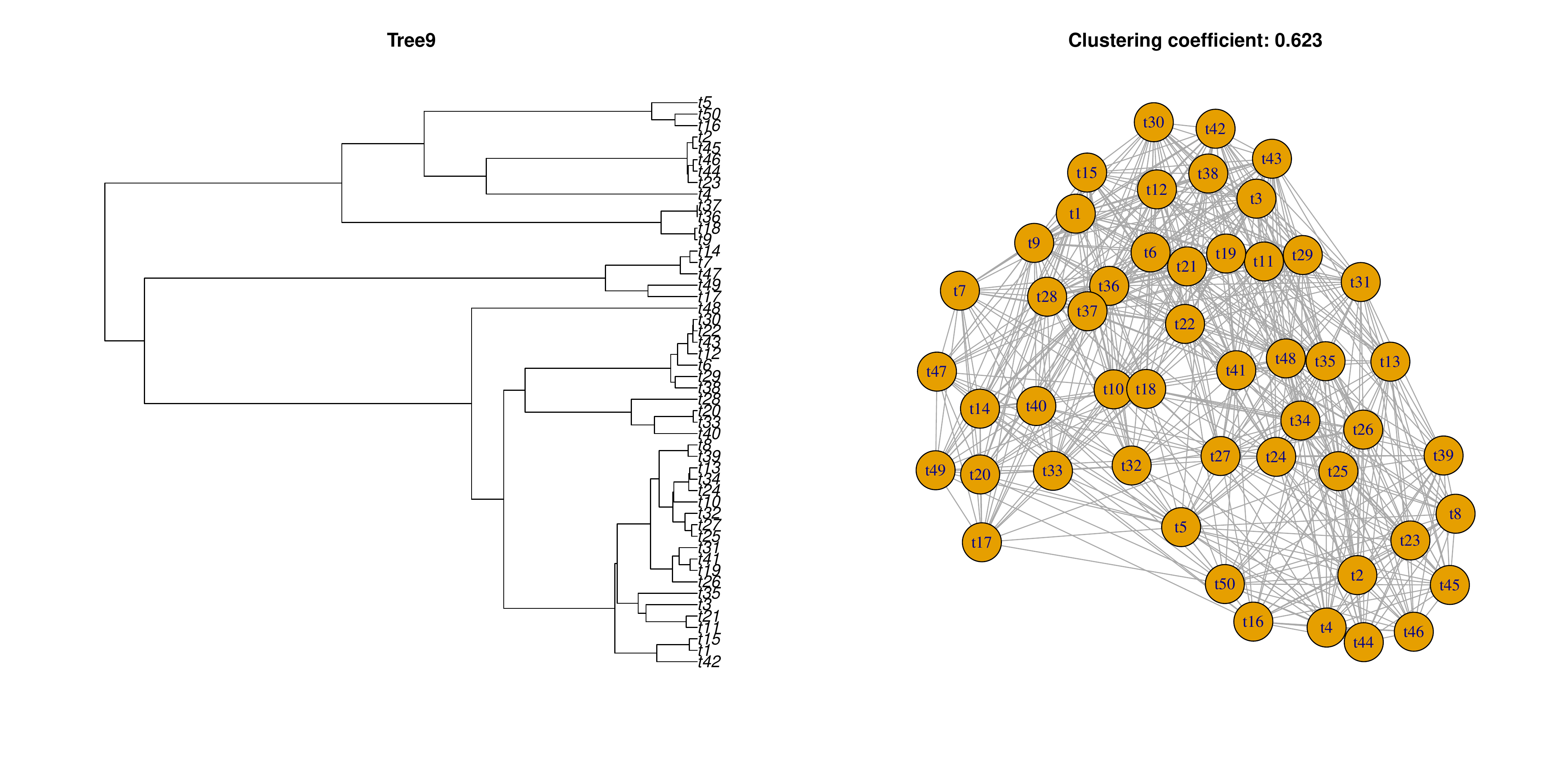}
	\includegraphics[width=1\textwidth]{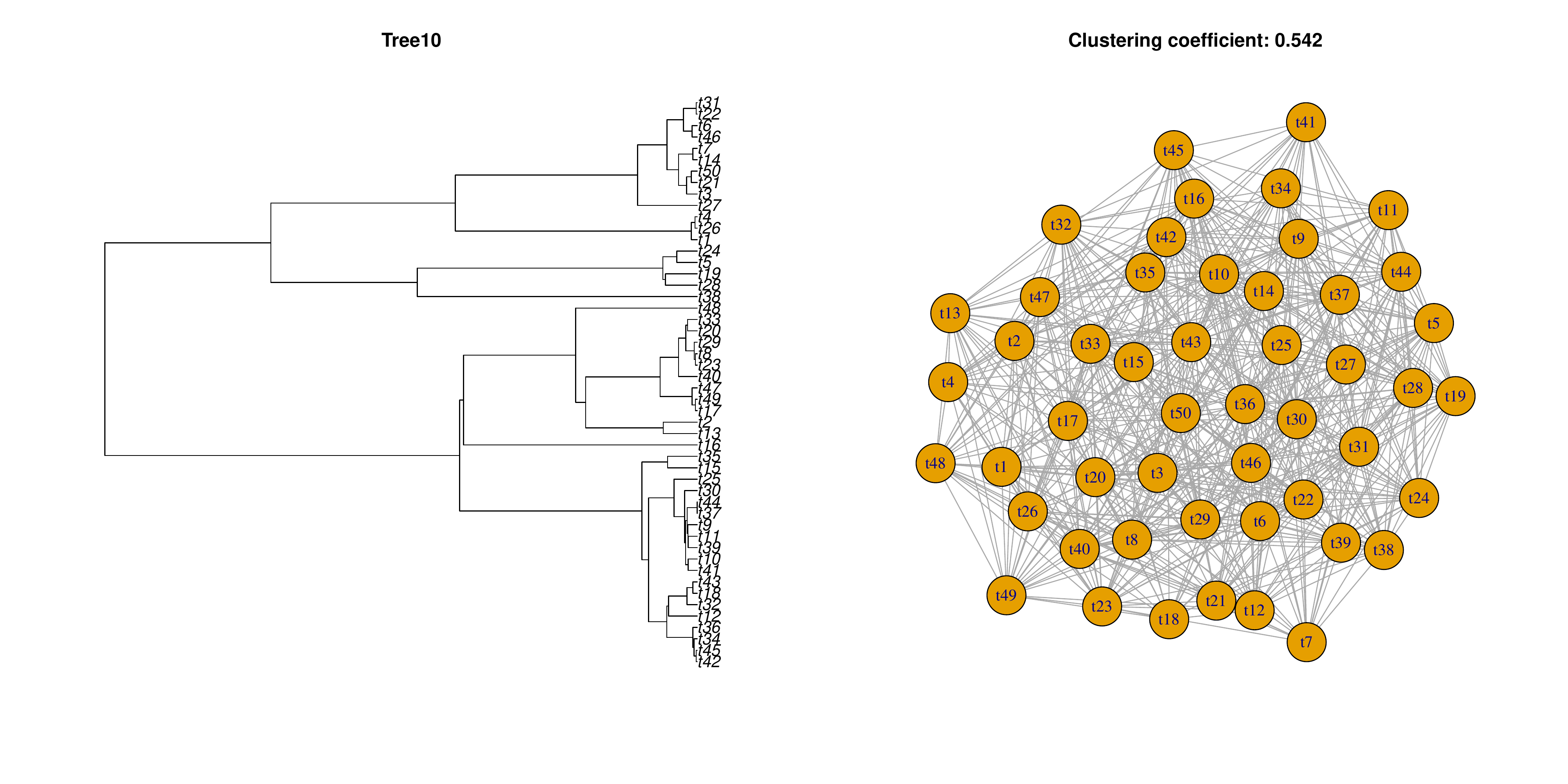}
	\caption{ $\Tcal_{9}$ and $\Tcal_{10}$ }\label{fig:graph910}
\end{figure}

\FloatBarrier

\bibliographystyle{asa}
\bibliography{PhyloBCG_Reference}
\end{document}